\documentclass[aps,reprint,superscriptaddress,prl,floatfix]{revtex4-2}
\usepackage[utf8]{inputenc}
\usepackage[T1]{fontenc}
\usepackage{amsmath,amssymb,graphicx,color,multirow}
\graphicspath{{figures/}{supplementary/figures/}}
\usepackage[bookmarks=false]{hyperref}
\usepackage[usenames, dvipsnames]{xcolor}
\usepackage{subcaption}
\usepackage[super]{nth}
\usepackage[normalem]{ulem}
\usepackage{booktabs}
\usepackage{xcolor}
\def\STO{SrTiO${}_3$~}

\def\LSNO{(La,Sr)NiO$_2$~}

\def\LNO{LaNiO${}_2$~}

\begin{document}

\title{Doping dependence of local moments in infinite layer nickelates}

\author{Martin Gonzalez}
\email{martin99@stanford.edu}
\affiliation{Stanford Institute for Materials and Energy Sciences, SLAC National Accelerator Laboratory, 2575 Sand Hill Road, Menlo Park, CA 94025, USA.}
\affiliation{Department of Materials Science and Engineering, Stanford University, Stanford, California 94305, USA.}

\author{Andreas Suter}
\affiliation{PSI Center for Neutron and Muon Sciences, 5232 Villigen PSI, Switzerland}

\author{Michal Kiaba}
\affiliation{Department of Materials Science and Engineering, Northwestern University, Evanston, IL, USA}

\author{Thomas Prokscha}
\affiliation{PSI Center for Neutron and Muon Sciences, 5232 Villigen PSI, Switzerland}

\author{Zaher Salman}
\affiliation{PSI Center for Neutron and Muon Sciences, 5232 Villigen PSI, Switzerland}

\author{Marc Gabay}
\affiliation{Laboratoire de Physique des Solides, Universit\'{e} Paris-Saclay, CNRS UMR 8502, F-91405 Orsay Cedex, France}

\author{Harold Y. Hwang}
\affiliation{Stanford Institute for Materials and Energy Sciences, SLAC National Accelerator Laboratory, 2575 Sand Hill Road, Menlo Park, CA 94025, USA.}
\affiliation{Department of Applied Physics, Stanford University, Stanford, CA 94305, USA.}

\author{Jennifer Fowlie}
\email{jfowlie@northwestern.edu}
\affiliation{Department of Materials Science and Engineering, Northwestern University, Evanston, IL, USA}

\date{\today}

\begin{abstract}
The infinite layer nickelates are notable for their lack of long-range antiferromagnetic ordering, in contrast to the parent compounds of the superconducting cuprates. Instead, the nickelates show evidence of short-range glassy behavior in both the undoped and optimally-doped regimes, implying that local electronic moments exist independent of superconductivity. However, the systematic doping-dependent magnetic behavior is not yet fully resolved, and characterizing it could uncover the relationship between local moments and the superconducting dome. In this work, we use muon spin rotation ($\mu$SR) on a (La,Sr)NiO$_2$ doping series from the undoped parent compound, through the superconducting dome, to the over-doped normal state (Sr substitution $0\% \leq x \leq 25\%$) to probe the magnetic ground state and the temperature-dependent static and dynamic behavior. We find that local moments experience spin freezing into a glassy state at temperatures on the order of a few tens of kelvin regardless of the doping level. We also observe a subtle destabilization of the glassy state with increased hole doping. These observations suggest that magnetism and superconductivity are largely decoupled phenomena with indirect interactions described in a multi-orbital framework.

%\textcolor{red}{This a first draft, feel free to add comments anywhere in this document. Sentences in red font are questions I have for anyone to clarify}
\end{abstract}
\pacs{NaN}

\maketitle

\section{Introduction}
%% General Intro on uncoventional SC in correlated systems
Since the discovery of high temperature superconductivity in the cuprates~\cite{bednorz1986possible}, much effort has been dedicated to understanding the origins of Cooper pairing in highly correlated systems. It has been widely believed that antiferromagnetic spin fluctuations play an important role in the pairing mechanisms that give rise to unconventional superconductivity~\cite{scalapino_common_2012,keimer_quantum_2015,plakida_spin-fluctuation_2001}. Therefore, studying the intrinsic magnetism of these correlated materials may uncover a common thread linking various classes of superconducting materials.

%% Intro to nickelates and distinction from cuprates
The realization of superconductivity in nickel-based compounds provides a novel materials system for developing a generalized understanding of unconventional superconductivity~\cite{li2019superconductivity}.  The infinite layer nickelates, $R$NiO$_2$ ($R$ = La, Pr, Nd,...), initially received attention due to their similarities with the cuprates, including a layered square-planar structure and nominal $3d^9$ configuration. However, experiment has revealed key differences in the electronic structures between these superconducting systems. Notably, the nickelates exhibit significant hybridization between the lanthanide $5d$ and Ni $3 d_{xz}/d_{yz}$, placing the understanding of nickelate superconductivity within a multi-orbital framework~\cite{hepting_electronic_2020, hepting2021soft,ding2024cuprate,sun2025electronic}. Computational work emphasizes the importance of this multi-orbital nature \cite{botana_similarities_2020, lechermann2020late}, including a self-doping effect.

%including the differing orbital alignment that established Mott-Hubbard character in the nickelates as opposed to the charge-transfer cuprates~\cite{goodge_doping_2021, gu_superconductivity_2022}.

In terms of magnetic properties, the cuprate parent compounds (e.g. La$_2$CuO$_4$) exhibit static ($\frac{1}{2}$,$\frac{1}{2}$) antiferromagnetic (AFM) order~\cite{vaknin1987antiferromagnetism} and the superconductivity upon doping is driven by Cu-$d_{x^2-y^2}$-wave pairing~\cite{tsuei2000pairing}. Upon hole-doping cuprates, the static AFM order is suppressed but paramagnon excitations with largely the same dispersion persist in the normal state ~\cite{le2011intense,dean_persistence_2013}.
By contrast, the parent compounds of the infinite layer nickelates show no evidence of long-range AFM order~\cite{hayward_synthesis_2003,ortiz_magnetic_2022}, although nickelates \textit{do} host ($\frac{1}{2}$,$\frac{1}{2}$) magnetic excitations reminiscent of $S = \frac{1}{2}$ magnons ~\cite{lu_magnetic_2021}. These paramagnon modes persist but are softened with increased hole doping. On the other hand, the homologous series of multilayer square planar neodymium nickelates, which do not require chemical doping, also exhibit magnetic excitations persisting across the superconducting dome and into the overdoped regime but with little doping dependence observed~\cite{pan2026superconducting}. The pairing symmetry of nickelates is still under discussion, but experiments indicate nodal structure~\cite{harvey2025evidence, ranna2025disorder}.

While these observations suggest a dominance of the Ni $d_{x^2-y^2}$ orbital, similar to cuprates, other experimental observations suggest that a complete description of the magnetic character of infinite layer nickelates necessitates invoking more than one orbital. Unlike cuprates, nickelate parent compounds exhibit short-range order with glassy behavior~\cite{ortiz_magnetic_2022,lin_universal_2022}. Recent studies also report spin glass behavior in Sr-doped superconducting nickelates~\cite{fowlie_intrinsic_2022,saykin_spin-glass_2025}. In some cases, lightly doped cuprates also exhibit a spin glass-like phase that bridges the undoped AFM order to the emergence of superconductivity~\cite{julien2003magnetic, aharony1988magnetic, hasselmann2004spin, tallon1997muon, stilp2013magnetic}. In cuprates, the onset of spin glass occurs below 30 K, while in nickelates the onset occurs between 100 and 200 K. The contrasting thermal energy scales, as well as the robustness of the nickelates' glassy behavior upon cooling through $T_c$, suggest that the unique multi-orbital nature of nickelates may provide multiple avenues for the distinct correlated electronic phases of superconductivity and magnetism, enabling their decoupling. %A detailed doping study examining the magnetic dynamics has not yet been carried out.

%The presence of similar magnetic excitations alongside markedly different magnetic ground states in the parent compound highlights the need to understand the interplay between magnetism and superconductivity in cuprates and nickelates. 
%Further work is needed to elucidate any doping-dependent change in the nickelates' magnetic behavior, and whether it is influenced by the effects of hole doping or superconductivity.

In conventional spin glass systems like binary alloy CuMn, the freezing temperature scales with the impurity concentration due to enhancement of the Ruderman-Kittel-Kasuya-Yosida (RKKY) interaction~\cite{binder1986spin, ruderman1954indirect}. In cuprates the glassy behavior is attributed to frustrated moments upon the introduction of holes to the CuO$_2$ planes. Consequently, hole doping initially raises the freezing temperature ($T_g$) until the system approaches the superconducting dome, at which point $T_g$ decreases ~\cite{kastner1998magnetic}. For the infinite layer nickelates a detailed doping study examining the magnetic dynamics has not yet been carried out.

In this work, we investigate the (La,Sr)NiO$_2$ doping series, with Sr substitution in the range $0\% \leq x \leq 25\%$, using low energy  muon spin rotation ($\mu$SR) to better understand the doping-dependent character of the magnetism. We focus on the La compounds, which have no 4\textit{f} moments, allowing us to isolate the role of Ni moments. We find that local moments persist through the doping series, with all samples exhibiting a gradual crossover from a room-temperature paramagnetic regime into a low-temperature state with full magnetic volume fraction. This crossover is characterized by the freezing of dynamic fluctuations at low temperatures as the system evolves towards a glassy state. These processes occur on similar thermal scales regardless of doping and we observe no anomalies at the onset of superconductivity -- either upon cooling through $T_c$ or upon hole-doping into or out of the superconducting dome. This reinforces the understanding that superconductivity and magnetism are largely decoupled in these systems and is consistent with the multi-orbital nature of infinite layer nickelates. Finally, our probes of dynamic magnetism suggest that there is a subtle doping dependence where an increase in hole density may destabilize the glassy state. We discuss this briefly in the context of multi-orbital physics.

\begin{figure*}[ht]
    \centering
    \includegraphics[width=\textwidth]{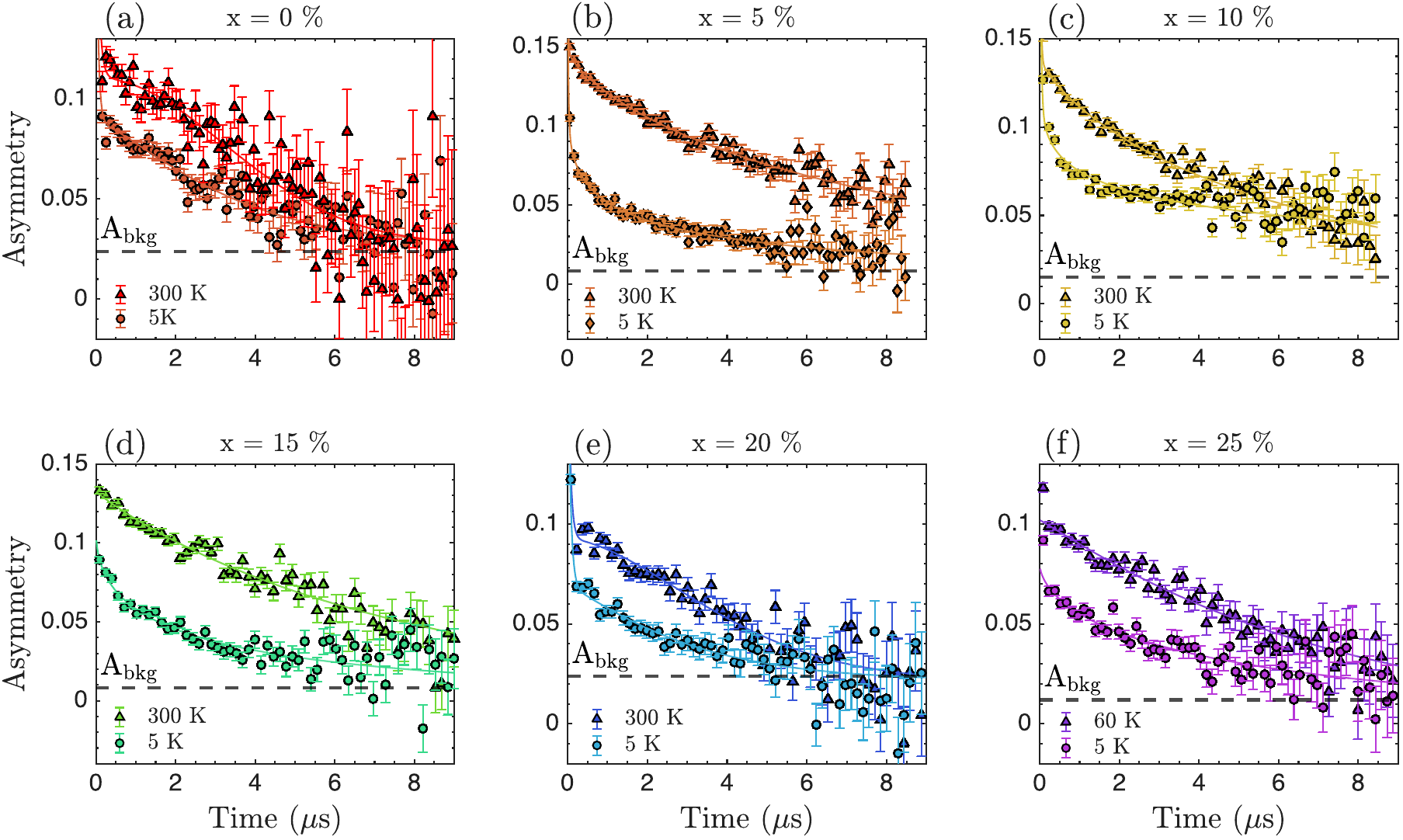}
    \caption{Zero-field (ZF) asymmetry spectra for the (La,Sr)NiO$_2$ nickelate doping series including (a) $x = 0\%$, (b) $x = 5\%$, (c) $x = 10\%$, (d) $x = 15\%$, (e) $x = 20\%$, and (f) $x = 25\%$. Each doping level shows the ZF spectra at $T = 300$ K and $T = 5$ K. Note that the $x = 25\%$ sample is the only exception, where the $T = 60$ K spectra is shown in lieu of the room temperature measurement where the data seems to be impacted by muon diffusion. The dashed line indicates the background asymmetry values derived from the nickel sample plate. The data are fitted with a stretched exponential function.
}
    \label{fig:ZF}
\end{figure*}

\section{Experimental}
\subsection{Sample Preparation}
We prepared a series of (La,Sr)NiO$_2$ thin films, first by growing the perovskite counterpart (La,Sr)NiO$_3$ as a precursor using pulsed laser deposition (PLD), followed by topochemical reduction with CaH$_2$ powder to induce a crystalline structural transformation into the infinite layer phase~\cite{wang2025molecular, yamamoto2013hydride}. The initial growth and subsequent topotactic reduction conditions used to prepare these samples are provided in the Supplementary Materials and discussed extensively elsewhere~\cite{lee2025effects, osada2023improvement}.  The samples consist of nickelate thin films capped with a SrTiO$_3$ layer, which serves to prevent reoxidation and improves the electronic transport properties~\cite{lee2020aspects, gonzalez2024absence}. Thin film pieces were organized into sample mosaics with at least 1 cm$^2$ total surface area for each representative doping level. Sample quality was verified using x-ray diffraction (XRD) and temperature-dependent resistivity measurements, $\rho$(T). 2$\theta$-$\omega$ symmetric scan fitting was used to confirm a complete transition into the infinite layer phase and to confirm the thicknesses for the sample heterostructures (See Figure~\ref{fig:Supp_XRD}). The electronic transport curves for each of the sample mosaics, as shown in Figure~\ref{fig:Supp_RT}, are consistent with previously reported nickelate samples across the hole-doping phase diagram~\cite{osada_nickelate_2021, lee_linear--temperature_2023}. Further details on the characterization of the infinite layer doping series are provided in the Supplementary Materials.

To maximize sensitivity to the nanometer-scale thickness of our films, we performed low-energy muon (LEM) experiments~\cite{morenzoni2002implantation,simoes_muon_2020, suter2023low} at the $\mu$E4 beamline of the Swiss Muon Source (S$\mu$S) at the Paul Scherrer Institut~\cite{prokscha2008new}. To maximize muon implantation in the sample, an incident beam energy of 3~keV was chosen to balance backscattering losses against the low thickness of the nickelate layer. Guided by Monte Carlo depth-profile simulations~\cite{eckstein2013computer}, the sample architecture was fine-tuned for this energy by optimizing the SrTiO$_3$ cap thickness and depositing an e-beam evaporated Au overlayer. Refer to  Figure~\ref{fig:Supp_hetero} for more details on sample heterostructure design.

% \begin{figure*}
%     \centering
%     \includegraphics[width=0.75\textwidth]{figures/Fig2_MVF_figure_v5.jpg}
%     \caption{Magnetic volume fraction (F$_M$) as a function of temperature for the \LSNOx nickelate doping series including (a) $x = 0\%$, (b) $x = 5\%$, (c) $x = 10\%$, (d) $x = 15\%$, (e) $x = 20\%$, and (f) $x = 25\%$. The dashed lines indicate the expected FM where the nickelate layer would become fully magnetic. Note that all samples approach a fully magnetic state at low temperatures, irrespective of the superconducting critical temperature, which is indicated by a vertical dashed line. Data in a) and e) are reproduced with permission from ref~\cite{fowlie_intrinsic_2022}. (g) Temperature- depolarization rate for the nickelate doping series, as measured using the weak transverse field (wTF) of 10 mT. (h) Temperature dependent depolarization for the $x = 5 \%$ measured at 10 mT (red) and 125 mT (orange).}
%     \label{fig:MVF}
% \end{figure*}

\begin{figure*}
    \centering
    \includegraphics[width=0.95\textwidth]{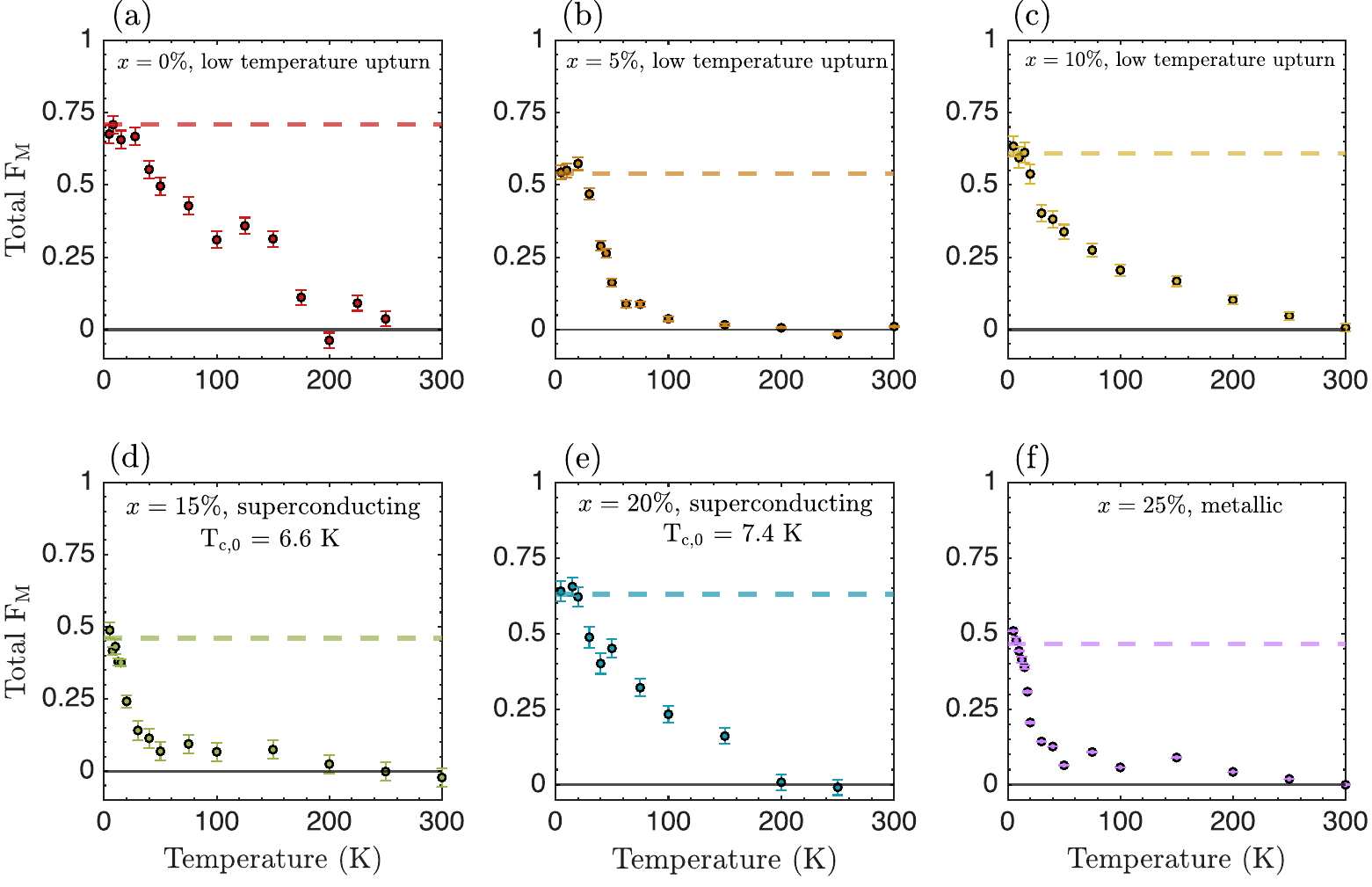}
    \caption{Magnetic volume fraction (F${\rm _M}$) as a function of temperature for the (La,Sr)NiO$_2$ series including Sr substitutions of (a) $x = 0\%$, (b) $x = 5\%$, (c) $x = 10\%$, (d) $x = 15\%$, (e) $x = 20\%$, and (f) $x = 25\%$. The dashed lines indicate the expected F$_{\rm  M}$ where the nickelate layer would become fully magnetic. Note that all samples approach a fully magnetic state at low temperatures, irrespective of the superconducting critical temperature. Data in a) and e) are reproduced with permission from ref~\cite{fowlie_intrinsic_2022}.}
    \label{fig:MVF}
\end{figure*}

\subsection{Principles of Muon Spin Rotation ($\mu$SR)}
Muon spin rotation spectroscopy, $\mu$SR,  is a technique that takes advantage of the muon's large magnetic moment to probe local magnetism within a material~\cite{blundell2021muon, le2011muon}. A spin-polarized beam of positively-charged muons is implanted into the sample, where each muon experiences Larmor precession at a frequency dependent on the local field, which can be the combined effects of the implantation site's nearby moments and any externally applied field. The implanted muons  subsequently decay and emit positrons preferentially in the direction of the muon spin~\cite{blundell2021muon}. By measuring the time-resolved angular distribution of emitted positrons, we obtain information regarding local moments. The number of positrons detected as a function of time is given by:

\begin{align}
    N(t) = N_0e^{-t/\tau_{\mu}}[1+A(t)]+ N_{\rm bkg}
    \label{eq:positron_detection}
\end{align}

where $N_0$ is a normalization factor setting an ``initial'' detection rate at $t=0$, $\tau_{\mu} =$ 2.2 $\mu s$ is the mean muon lifetime, and N$_{\rm bkg}$ refers to uncorrelated background events. The positron counts are used to determine the asymmetry function, $A(t)$, which is defined to correlate with the time between a single muon entering the experiment and the respective positron being detected. The asymmetry is often directly called “$\mu$SR signal”. The information gained from a $\mu$SR measurement depends on the configured experimental geometry. The measurement geometries used in this work are zero-field (ZF), weak-transverse field (wTF) and longitudinal field (LF) configurations~\cite{amato2024,blundell2021muon}. The distinction between these geometries is the relative orientation between the incident muon spin polarization and the applied magnetic field. In dynamical magnetic systems, we can probe electronic moments that have correlation times that are longer than around 1 $\mu s$, i.e., are static on the muon timescale. The specifics of these measurement geometries will be discussed further in the Results Section.  

\section{Results}
\subsection{Probing Electronic Moments with ZF $\mu$SR}
The ZF configuration applies no external field, meaning that any field experienced by the muon spin ensemble originates from moments within the sample. Hence, the ZF measurement is a sensitive probe of even weak internal magnetism and the ZF asymmetry spectrum can be used to identify the presence of magnetic moments and distinguish between their different origins. Because nuclear moments are temperature-independent, our low-temperature investigation isolates the role of Ni electronic moments. Figure~\ref{fig:ZF} shows the measured ZF spectra for the nickelate doping series for limited temperatures, while Figure~\ref{fig:Supp_ZF} displays all the temperatures recorded.

The ZF spectra provide evidence for the presence of local electronic moments that are quasi-static on the muon timescale within the measured temperature range between T = 300 K and T = 5 K.  The evolution in the functional form of the asymmetry curve between the room temperature and base temperature measurements suggests the emergence of a distinct magnetic ground state at low temperatures. An exception to the measurement parameters is the $x = 25\%$ sample where the measurement at T = 300 K showed evidence of muon diffusion, which is also observed below room temperature in cuprates~\cite{sonier2002correlations, pal2018quasistatic}. Hence, for that sample only, we show the ZF asymmetry at T = 60 K for a comparison of the evolving magnetic state before the thermally-activated diffusion onsets. A brief discussion on muon diffusion, can be found in the Supplementary Materials. 

For all doping levels, the ZF spectra exhibit an increasingly damped exponential form upon cooling. This damping is indicative of the emergence of local magnetism resulting from the freezing of electronic moments. This general trend is observed across the entire doping series and is consistent with prior ZF measurements on undoped powder LaNiO$_2$ samples~\cite{ortiz_magnetic_2022}. In addition, the asymmetry curves feature a tail that diminishes over longer timescales, indicative of spin dynamics in this system~\cite{fowlie2023metal}. The ZF spectra can be modeled using depolarization functions specifically derived to describe a given type of magnetism.  For instance, the Kubo-Toyabe function is a Gaussian-like curve that describes the muon response to nuclear moments, i.e., the absence of electronic moments~\cite{fowlie_intrinsic_2022, kiaba_observation_2024, amato2024, hayano1979zero}. Alternatively, the ZF spectra can be modeled with a more general stretched exponential, $A^{\rm ZF} = A^{\rm ZF}_0e^{-(\lambda_{\rm ZF}  t)^\beta}$. The resulting stretched exponential fits are displayed as solid lines in Figure~\ref{fig:ZF}. The advantage in using the stretched exponential function is its flexibility in accounting for a wide variety of magnetic systems. In particular, stretched exponential fits can capture general trends associated with magnetically disordered and inhomogeneous systems~\cite{blundell1999spin}. For instance, the temperature-dependent shape of the stretched exponential, as represented by its lineshape parameter $\beta$, can be indicative of broad relaxation phenomena attributed to spin glass behavior~\cite{campbell1994dynamics}. 
See the Supplementary Material for a more complete discussion of the stretched exponential fit (Fig.~\ref{fig:Supp_Beta}) and trends in its extracted parameters~\cite{suter2012musrfit}. For all ZF curves we find $\beta < 2$, indicating the presence of electronic moments over the entire doping series. $\beta$ also tends to decrease gradually towards lower temperature, indicating a crossover to a magnetic ground state in (La,Sr)NiO$_2$ that may be short-range ordered or glassy in nature.

\subsection{Revealing fully magnetic nickelates with wTF $\mu$SR}

The wTF geometry can be used to determine the fraction of the sample volume occupied by a magnetic phase. A weak field is applied perpendicular to the initial muon spin orientation, and the resultant asymmetry depends on the relative strength and distribution of the sample's internal and applied fields. In the limiting case where the applied field dominates, muons will precess at a Larmour frequency corresponding to the external field, maximizing the amplitude of the Larmor precession signal. This limiting case is applicable when the material is diamagnetic or paramagnetic. However, if the internal fields become comparable to the external field, the spins of the muon ensemble will quickly dephase. Towards low temperature, the emergence of electronic moments that are static on the muon timescale diminishes the wTF asymmetry. The asymmetry oscillations arising from muon precession under a transverse field, B, is given by:
\begin{align}
    A(t) = A_0e^{-\lambda_{\rm TF} t}~\cos{(\gamma_{\mu}Bt + \phi)}
    \label{eq:wTF_oscillations}
\end{align}
where $A_0$ is the initial asymmetry amplitude, $\lambda_{\rm TF}$ is the transverse-field depolarization rate, $\gamma_{\mu}$ is the muon gyromagnetic ratio, and $\phi$ is a relative phase factor with respect to the initial spin orientation. The magnetic volume fraction, $F_{\rm M}$, is calculated from:

\begin{align}
    F_{\rm M}(T) = 1 - \frac{A_0(T)}{A_{\rm PM}} 
    \label{eq:MVF}
\end{align}

For each doping level, $A_{\rm PM}$ is taken from the room temperature asymmetry, where the samples are assumed to be paramagnetic, and zero magnetic volume fraction is defined. $F_{\rm M}(T)$ for the doping series is shown in Figure~\ref{fig:MVF}. In this work, the total $F_{\rm M}$ refers to the volume of the overall sample heterostructure (including SrTiO$_3$ and Au capping layers) that becomes magnetic. The horizontal dashed line represents the threshold $F_{\rm M}$ value corresponding to a fully magnetic nickelate layer, as determined by the fraction of incident muons implanted in the film calculated using TRIMP.SP. We find that the nickelate samples reach a fully magnetic state across all doping levels at temperatures between 5 - 15 K. The steepest increase in $F_{\rm M}$ towards a fully magnetic state occurs at around $T = 40$ K, indicating a crossover towards a static (on the timescale of the muon) magnetic ground state. The persistence of a fully magnetic state across all samples indicates that magnetism is intrinsic, i.e., cannot be attributed to defects, and occurs independently of the hole concentration. Notably, the behavior of the superconducting samples ($x = 15\%$ and $x = 20\%$ with respective $T_{c,0}$ values of $6.6$ K and $7.4$ K) is qualitatively the same as the non-superconducting compositions. The persistence of a fully magnetic state stands in contrast with other superconducting systems, including the cuprates and iron pnictides, which exhibit a suppression of magnetic volume fraction with increased doping~\cite{bernhard2012muon, takeshita2009competition}. We suggest that magnetism, in a broad sense, is decoupled from superconductivity in infinite layer nickelates.

% This theoretical $F_{\rm M}$ subtracts asymmetry contributions from the SrTiO$_3$ substrate~\cite{salman2014direct} and from the background asymmetry of the sample plate and differs from sample to sample due to the variation in the thicknesses of the nickelate and capping layers. The muon implantation profile were estimated using Monte Carlo simulations \cite{eckstein2013computer}, allowing us to engineer a sample heterostructure that maximizes muon implantation into the nickelate layer.

\begin{figure}
    \centering
    \includegraphics[width=\columnwidth]{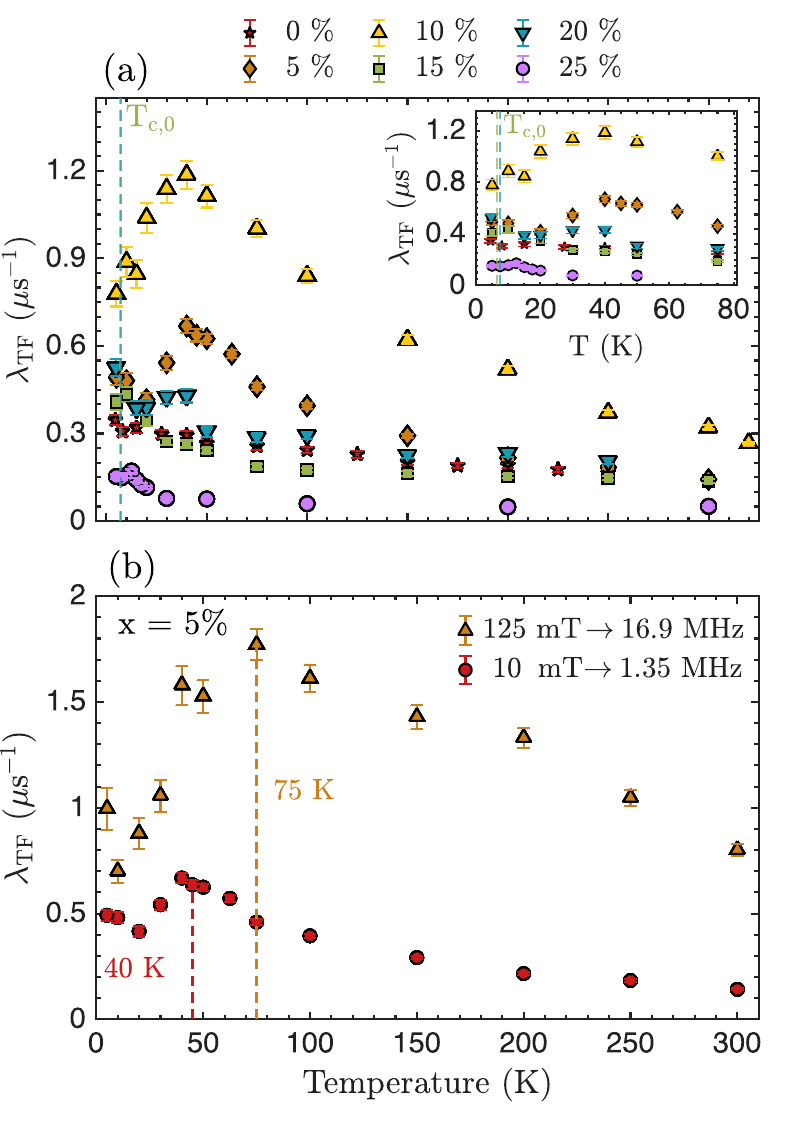}
    \caption{(a) Temperature-dependent depolarization rate for the (La,Sr)NiO$_2$ series, measured using a weak transverse field (wTF) of 10 mT. The inset is an enlarged view at low temperatures of $\lambda_{\rm TF}$(\textit{T}). (b) Temperature dependent depolarization for the $x = 5 \%$ sample with measured fields corresponding to Larmor frequencies of 1.35 MHz (red) and 16.9 MHz (orange).}
    \label{fig:depol_rate}
\end{figure}

% While the magnetic volume fraction establishes that the intrinsic magnetism is independent of superconductivity, no information is revealed of any doping dependence on the dynamics of local moments. Therefore, we also use the wTF geometry to measure the temperature dependent depolarization rate for each doping level. \textcolor{red}{I've seen in the literature the depolarization rate be plotted using different geometries i.e ZF orientation in ref~\cite{ortiz_magnetic_2022}. Do the depolarization rate in different geometries tell us different things?}%  
% The depolarization rate describes the rate at which the muon beam's polarization is lost in its magnetic environment. In transverse-field measurements, the depolarization rate is dominated by the loss of coherent muon precession due to field inhomogeneity or the emergence of magnetically ordered regions, and is therefore closely related to the magnetic volume fraction. 

Fitting the asymmetry oscillations from Equation~\ref{eq:wTF_oscillations} also allows extraction of the temperature-dependent depolarization rate, $\lambda_{\rm TF}$(T), which is shown in Figure~\ref{fig:depol_rate}(a). We see a broad peak at 40 K in the underdoped samples ($x = 5\%,~ x = 10\%$) and another weaker peak at 15 K in the overdoped sample ($x = 25\%$). Peaks in the $\mu$SR depolarization rate indicate that, upon cooling, magnetic moments begin to appear static on the measurement timescale. The depolarization rate is largest when the spin fluctuation rate becomes comparable to the Larmour precession frequency set by the applied field. A broad $\lambda_{\rm TF}$ peak is commonly linked to glassy freezing, where moments slow down and begin to freeze in a disordered and gradual manner over a wide temperature range~\cite{murnick1976muon, emmerich1981mu+, brown1981anomalous}. At lower temperatures, as the spins become predominantly frozen, the depolarization rate decreases because contributions from dynamic fluctuations are reduced, even though static magnetism continues to grow. The observation of dynamical peaks in the $\mu$SR depolarization rate is consistent with recent $^{17}$O NMR studies on LaNiO$_2$ that report a similar broad peak in the spin-lattice relaxation rate ($1/T_1$) at $T \approx 52$ K~\cite{zhou2025origin}. The overall temperature-dependence in $1/T_1$ is further evidence of the freezing of local spins and the system's approach to spin-glass-like behavior at lower temperatures.

The peak in $\lambda_{\rm TF}$ changes across the doping series, appearing to evolve from 40 K to 15 K as the doping is increased to x = $25\%$. Similarly, $^{17}$O NMR spectroscopy has shown a shift in the relaxation rate peak to lower temperatures with the introduction of 18 \% Sr doping~\cite{zhou2025origin}. However, our other samples ($x = 0\%,~ x = 15\%,~ x = 20\%$) have no clear peak but rather show a modest increase in depolarization at low temperatures. If spin freezing occurs over a wide enough distribution of temperatures, $\lambda_{\rm TF}$ will show a gradual increase as static magnetism develops. In this scenario, different spatial regions of the sample can freeze at different temperatures which is observed in spin glass systems~\cite{lin_universal_2022}. The lack of a distinct peak in $\lambda_{\rm TF}$ at these doping levels, despite our evidence for intrinsic magnetism, reveals the subtle effects of hole-doping on the nickelates' magnetic properties. For instance, ref.~\cite{zhou2025origin} posits that introducing holes could alter the activation energy for spin
freezing.  Overall, the emergence of intrinsic magnetism remains robust across the series, but the underlying magnetic interactions and dynamics may be sensitive to the doping concentration. 

\begin{figure}
    \centering
    \includegraphics[width=\columnwidth]{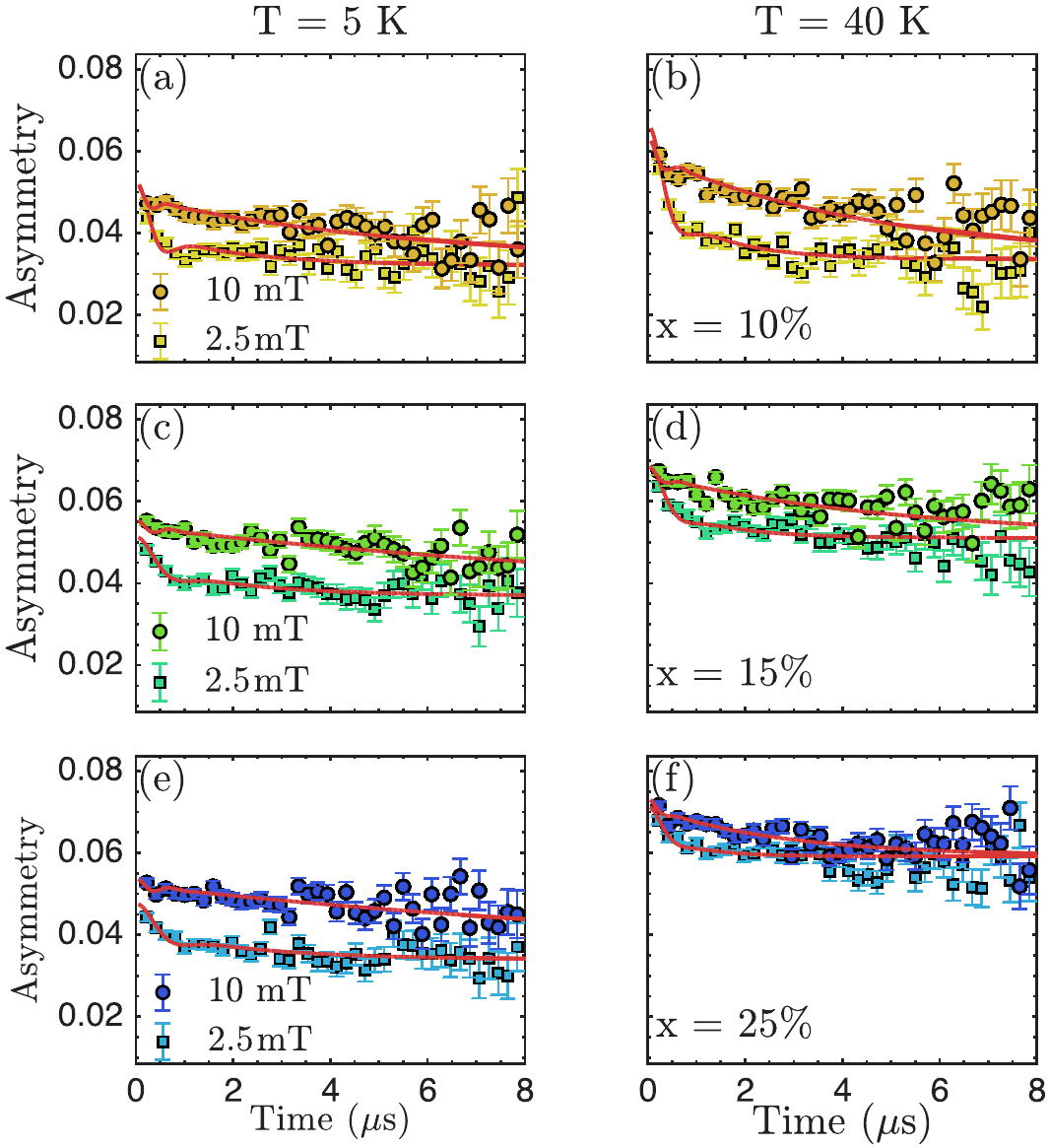}
    \caption{Longitudinal field spectra of (La,Sr)NiO$_2$ samples at (a-b) $x = 10\%$, (c-d) $x = 15\%$, and (e-f) $x = 25\%$ with applied fields at magnitudes of 25 G and 100 G. The measurements were taken at T = 5 K (a, c, e) and T = 40 K (b, d, f). The solid red lines indicate global fits of the LF spectra. }
    \label{fig:LF}
\end{figure}

Previous works have identified a short-ranged ordered spin glass via magneto-optic effects and with AC susceptibility measurements on bulk powder samples~\cite{saykin_spin-glass_2025, ortiz_magnetic_2022, lin_universal_2022}. Using the TF-$\mu$SR geometry, one can vary the strength of the applied field to tune the Larmor precession frequency to conduct an experiment that is similar to a frequency-dependent AC susceptibility. In general, $\lambda_{\rm TF}$ is comprised of contributions from both static inhomogeneous field distributions and dynamic spin-lattice relaxation~\cite{sonier2010high}. However, static inhomogeneous broadening does not exhibit an intrinsic frequency-dependent temperature shift, therefore peak shifts are dominated by dynamic fluctuations. As shown in Fig.~\ref{fig:depol_rate}(b), the peak depolarization rate for the x = 5\% sample shifts from 40 K to 75 K as the Larmor frequency is increased from $f = 1.35 $~MHz to $f = 16.9$~ MHz. A similar frequency-dependent shift in relaxation rate has been observed in under-doped La$_{2-x}$Sr$_x$CuO$_4$ cuprates using LF $\mu$SR. The peak in longitudinal depolarization rate shifts upward in temperature at higher fields (hence higher frequency) indicative of a spin glass state \cite{borsa1995staggered}.    In both cases, probing at higher Larmor frequencies shifts the condition for maximum relaxation ($\omega \tau_c \approx 1$) toward higher temperatures where spin correlation times $\tau_c$ are shorter. Hence our $\mu$SR data is consistent with previous findings in the literature, and, by extension, establishes the persistence of the glassy behavior over all measured doping levels.

Our $\mu$SR wTF data has verified a fully magnetic ground state intrinsic to the nickelates across the doping series. While the magnetism persists independent of superconductivity, we have also disentangled a more subtle doping dependence of the dynamics of the electronic moments. To further explore the effect of doping on the nickelate ground state in the dynamic regime, we use LF measurement geometry.

\subsection{Identifying Dynamic Fluctuations using LF $\mu$SR}
As mentioned previously, the ZF configuration is useful in identifying the presence of local moments within a sample. However this experimental geometry is unable to differentiate between broad distributions of static moments and dynamic fluctuating moments. Both scenarios can produce an exponential decay shape in the ZF asymmetry loss spectra, although the long time decay towards zero ZF asymmetry suggests residual dynamics. To distinguish static disordered moments from dynamic fluctuating moments, we use the longitudinal field (LF) geometry which applies an external field parallel to the initial spin polarization of the incident muon beam, inducing Zeeman splitting of the muon spin states. Under a sufficiently strong longitudinal field, a sample with static moments will not have muon precession due to the quenching of any transverse component from internal fields. In this scenario, muons maintain a stationary Zeeman state, and the implanted muons will not depolarize, resulting in a time-independent, i.e., flat, asymmetry within $\mu$SR timescales.  However, a sample that hosts dynamic fluctuations will drive spin-flip transitions between energy split Zeeman levels, resulting in an asymmetry curve that decays with time. By measuring the asymmetry loss response under a longitudinal field, we can verify the presence of dynamical fluctuations. Therefore, we measured the LF spectra of films representative of the under-doped ($x = 5\%$), superconducting ($x = 15\%$), and over-doped ($x = 25\%$) regimes. We conducted these measurements at $T = 40$ K, the temperature showing the most rapid approach towards 100\% magnetic volume fraction, and at  $T = 5$ K where we would expect frozen spins. Results are displayed in Figure~\ref{fig:LF}. 

\begin{figure*}
    \centering
\includegraphics{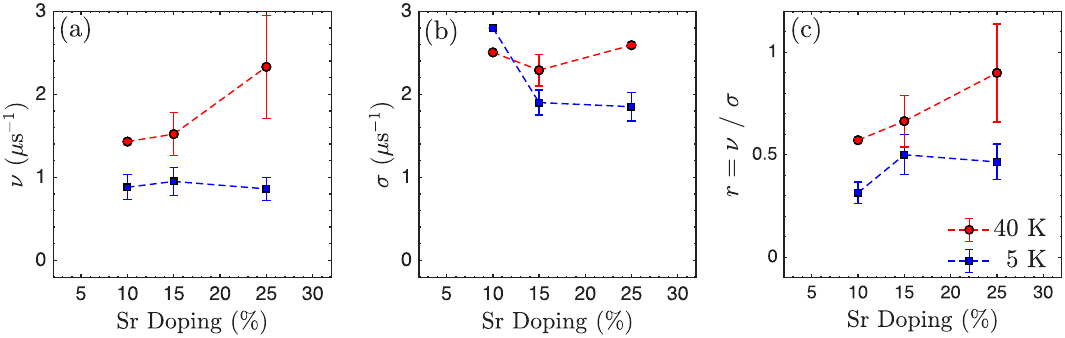}
    \caption{Doping-dependent parameters extracted from global fitting of the LF spectra. Fit parameters include (a) the fluctuation rate $\nu$, (b) the internal field width $\sigma$, and (c) the ratio of these two quantities. }
    \label{fig:LF_fits}
\end{figure*}

For all doping levels, the LF spectra measured at 40 K show a decaying asymmetry profile at both 25 and 100 G, this behavior is indicative of dynamic moments. The LF spectra at 5 K show similar behavior, although the asymmetry is significantly flattened at 100 G, which suggests slower fluctuations. To better quantify the strength of fluctuations between these two temperatures, the data was fitted using a dynamic Gaussian Kubo-Toyabe function, whose form is shown in the Supplementary Materials~\cite{hayano1979zero, de1992quantum, keren1994generalization}.  The solid red lines indicate fits to the LF spectra and illustrate the overall shape of the asymmetry data.  From the fitting function we extract $\nu$, the fluctuation rate of dynamic moments, and $\sigma$ which is related to the second moment of the field distribution by $\sigma/\gamma_{\mu} = \sqrt{\langle\Delta B^2\rangle}$. These extracted parameters are plotted against doping in Figure~\ref{fig:LF_fits}. 
 
The fluctuation rate, as extracted from the global fit curves, is at least 1.5$\times$ greater at 40 K compared to 5 K, consistent with the slowing down of dynamic moments as the sample is cooled. Furthermore, we see that the low-temperature fluctuation rate is relatively constant across all doping, meaning that the slow spin dynamics are not sensitive to Sr-doping at 5 K. In contrast, the fluctuation rate increases with doping when measured at 40 K. This trend is counter to conventional dilute spin glasses, where impurities slow down dynamics and increase the freezing temperature, hinting at more complex magnetic interactions induced by the introduction of holes. 
%Nevertheless, the relative insensitivity of the dynamics to Sr-doping at lower temperatures corroborates our physical interpretation of fluctuating local moments freezing towards a spin glass state.  % \textcolor{red}{What can we conclude based on the trends given here? Its un-intuitive why the field width would decrease at a higher fluctuation rate}

While the doping dependence of $\sigma$, roughly the field distribution, is difficult to interpret, this parameter is important in quantifying the relative importance of static and dynamic contributions to the asymmetry spectra. The ratio, $r = \nu$ / $\sigma$ (Figure~\ref{fig:LF_fits}(c)) is an indicator of whether the magnetic moments are mostly dynamic ($r\gg$ 1) or static ($r\ll$ 1) on the timescale of $\mu$SR measurements. For magnetic systems that exist within either of these limiting cases, the ZF asymmetry spectra can usually be analyzed using a product function consisting of a static Kubo-Toyabe relaxation component, and a dynamical component captured by an exponential function~\cite{ryan2004zero}. Within the temperature range 5 K < T < 40 K, the infinite layer nicketales exhibit an intermediate regime, $r \sim$ 1, where both static and dynamic contributions are relevant on measurement timescales. In the intermediate regime, the product function therefore cannot be used to fit the ZF asymmetry~\cite{ryan2004zero}, justifying our use of a stretched exponential function in our earlier analysis. However, we note that $r$, which is already less than 1, decreases even more from 40 K to 5 K, showing a growing static contribution from spin freezing at lower temperatures. 

\section{Discussion}
In this study we used $\mu$SR to investigate the doping-dependence of the intrinsic magnetism of superconducting infinite layer lanthanum nickelates. While previous studies have identified a spin glass state in the undoped and optimally-doped nickelates, an investigation of the dynamics of local electronic moments across the hole-doping phase diagram, including over- and under-doped compounds, was previously absent. We find that the emergence of intrinsic local magnetic moments occurs regardless of the doping level ($0\% \leq x \leq 25\%$). By extension, the local moments are insensitive to superconductivity, as we observe no anomalies in either ZF or wTF measurements at the superconducting transition temperature. 

In general, while the presence of defects could theoretically be responsible for the emergence of magnetism, the defect density would have to be almost one defect per unit cell to account for the observed 100\% low-temperature magnetic volume fraction. Atomic resolution imaging of similar samples show limited defects, meaning that the observed magnetism is likely intrinsic~\cite{osada_nickelate_2021}.

Further aspects of the nature of the intrinsic local magnetism are revealed by investigating the temperature-dependent spin dynamics of the system. The observation of broad peaks in the wTF depolarization rate reveal a wide distribution of spin relaxation timescales upon cooling. The temperature of the peak, when present, is lowered with increased hole doping, suggesting faster fluctuations and less thermal energy required to melt the moments.

Associated wTF field-dependent measurements served as direct evidence of low-temperature glassy behavior. This finding corroborates our prior work using magneto-optical Kerr effect on similar nickelate samples grown on the same SrTiO$_3$ substrates that we use here ~\cite{saykin_spin-glass_2025}. Based on the magnetic volume fraction, spin freezing appears to set in at \textasciitilde{~}40 K, which is in close agreement with the onset temperature of the optical behavior attributed to spin glass ~\cite{saykin_spin-glass_2025}. In both experiments, the onset temperature of the glassy behavior is not impacted by whether or not there is a superconducting transition, further demonstrating a decoupling of superconductivity and magnetism. We also note that the onset temperature for glassy behavior coincides with the temperature showing sign reversal in the hall coefficient; and where an insulating upturn is observed in under-doped samples~\cite{Osada_Hall_LSNO_2025}. This may suggest a global change in the Fermi surface topology that drives changes in both scattering channels and local moments.

We further verified gradual spin freezing towards a low-temperature glassy state with longitudinal field (LF) measurements, which showed the slowing down of dynamic fluctuations below 40 K. Similar to the TF measurements, we observed a slight increase in fluctuation rate with hole doping at 40 K, possibly revealing subtle doping-dependent changes in magnetic interactions and spin dynamics. 

We now propose some plausible electronic frameworks in which to interpret our observations of (1) an apparent decoupling between superconductivity and magnetism and (2) a subtle increase in fluctuation rate with increased hole doping. We also discuss other reported experiments.

First, the coexistence of magnetism and superconductivity suggests that local moments only weakly interact with the delocalized electrons that form cooper pairs. This phenomenon may be understood within a multi-orbital framework of superconductivity in the infinite layer nickelates. It has been established that chemical substitution of the lanthanide sites with divalent cations produce holes that reside within the in-plane Ni 3$d_{x^2-y^2}$ orbitals, which are believed to be responsible for superconductivity~\cite{goodge_doping_2021, rossi2021orbital}. The momentum-resolved electronic structure of optimally-doped nickelates highlights the importance of these in-plane orbitals, with angle-resolved photoemission measurements revealing robust Fermi surface sheets derived from highly-correlated Ni $d_{x^2-y^2}$ bands~\cite{sun2025electronic}.  Additionally, the Fermi surface features itinerant electron pockets at the Brillouin zone corners that originate from the hybridization of the out-of-plane Ni orbitals with the lanthanide 5$d$ orbitals. The highly dispersive nature of these electron pocket states makes them unlikely hosts for localized spins. Therefore, as a possible explanation for observation (1), we surmise that these pockets instead mediate an RKKY coupling between local moments residing on the highly correlated, in-plane Ni 3\textit{d} orbitals.

The orbital and spin character of the doped holes factors into the magnetic interactions dictating the doping-dependent glassy behavior. Depending on the relative strengths of the Ni 3\textit{d} crystal field splitting and Hund's coupling, the doped holes could either adopt a high-spin (HS) S = 1 triplet state or a low-spin S = 0 (LS) singlet state, deviating from the undoped S = $\frac{1}{2}$ background~\cite{lechermann2020multiorbital}. Rossi et al. have argued in favor of a LS configuration for doped holes in the Nd compounds~\cite{rossi2021orbital}.  In cuprates it is believed to be the S = 0 doped hole singlet state that is simultaneously responsible for the suppression of antiferromagnetism and the onset of superconductivity~\cite{zhang1988effective}. While our measurements cannot distinguish between HS and LS configurations with certainty, our overall observation of a decoupling of local moments and superconductivity tends to favor the HS picture. Regardless, in either the LS or HS configuration, the magnetic impurity can be taken as $\Delta$S = $-\frac{1}{2}$ or $\Delta$S = $\frac{1}{2}$ respectively and an RKKY interaction via the itinerant electron pocket can be described.

We now consider our observation (2), which is in contrast to conventional spin glasses where increasing magnetic impurities increase the freezing temperature~\cite{cannella1972magnetic, mulder1981susceptibility}. Regardless of the spin character of the doped holes, we expect an increasing density of magnetic impurities with greater Sr-doping. The subtle trend we observe with hole doping - that the fluctuation rate increases and the onset freezing temperature decreases - can therefore only be reconciled if the RKKY interaction strength itself is decreasing with doping. This is plausible if the itinerant 5\textit{d}/4\textit{f} hybridized pockets are the mediators and their occupation diminishes with doping.

How the 5\textit{d}/4\textit{f} bands evolve with doping is not certain and may depend on the lanthanide ion. Recent optical conductivity measurements have probed the doping-dependent electronic structure of Nd-based infinite layer nickelates, and have found that increased hole doping depletes, but does not completely eliminate, the electron pocket, even well into the overdoped regime, suggesting a deviation from a rigid band shift picture~\cite{kim2026optical}. This is consistent with our observation that intrinsic magnetism is established across the superconducting dome but that the interaction strength may be subtly weakened with doping. On the other hand, recent x-ray absorption spectroscopy experiments on the La compounds, analogous to those we study here, demonstrate that the effect of hole-doping is to shift spectral weight from the Ni 3\textit{d} orbitals to hybridized O 2\textit{p} orbitals and that the holes have little effect on the La 5\textit{d}/4\textit{f} hybridized band occupation~\cite{tang2026ni}. However, they find a limited linear dichroism, suggesting that the multi-orbital picture in these compounds may also include mixing of the Ni in-plane and out-of-plane orbitals. In this scenario, the RKKY interaction may still be weakened if the shifting covalency also depletes the Ni out-of-plane orbitals via interorbital mixing. How the lanthanide ion affects the complex interplay between magnetism, superconductivity, and doping in infinite layer nickelates is unknown and worthy of further study.

In conclusion, our doping-dependent $\mu$SR measurements demonstrate that the presence of local electronic moments and low-temperature glassy spin freezing persist across the (La,Sr)NiO$_2$ phase diagram for Sr substitutions of ($0\% \le x \le 25\%$), remaining fundamentally agnostic to the emergence of superconductivity. The observation of a fully magnetic phase across both non-superconducting and superconducting regimes confirms that local-moment magnetism is an intrinsic property of infinite-layer nickelates. These findings point towards a multi-orbital electronic framework that allows magnetism and superconductivity to coexist.

\subsection{Acknowledgments}
The work conducted at Stanford/SLAC was supported by the U. S. Department of Energy, Office of Basic Energy Sciences, Division of Materials Sciences and Engineering (Contract No. DE-AC02-76SF00515). Part of this work is based on experiments performed at the Swiss Muon Source S$\mu$S, Paul Scherrer Institute, Villigen, Switzerland. 

\bibliographystyle{apsrev4-2}
\bibliography{bib/muSR_citations}

@article{botana_similarities_2020,
	title = {Similarities and {Differences} between {LaNiO}$_{\textrm{2}}$ and {CaCuO}$_{\textrm{2}}$ and {Implications} for {Superconductivity}},
	volume = {10},
	issn = {2160-3308},
	url = {https://link.aps.org/doi/10.1103/PhysRevX.10.011024},
	doi = {10.1103/PhysRevX.10.011024},
	language = {en},
	number = {1},
	urldate = {2022-04-11},
	journal = {Physical Review X},
	author = {Botana, A. S. and Norman, M. R.},
	month = feb,
	year = {2020},
	pages = {011024},
}

@article{fowlie_intrinsic_2022,
 title={Intrinsic magnetism in superconducting infinite-layer nickelates},
  author={Fowlie, Jennifer and Hadjimichael, Marios and Martins, Maria M and Li, Danfeng and Osada, Motoki and Wang, Bai Yang and Lee, Kyuho and Lee, Yonghun and Salman, Zaher and Prokscha, Thomas and others},
  journal={Nature Physics},
  volume={18},
  number={9},
  pages={1043--1047},
  year={2022},
  publisher={Nature Publishing Group UK London}
}

@article{osada_nickelate_2021,
	title = {Nickelate {Superconductivity} without {Rare}‐{Earth} {Magnetism}: ({La},{Sr}){NiO} $_{\textrm{2}}$},
	volume = {33},
	issn = {0935-9648, 1521-4095},
	shorttitle = {Nickelate {Superconductivity} without {Rare}‐{Earth} {Magnetism}},
	url = {https://onlinelibrary.wiley.com/doi/10.1002/adma.202104083},
	doi = {10.1002/adma.202104083},
	language = {en},
	number = {45},
	urldate = {2023-03-02},
	journal = {Advanced Materials},
	author = {Osada, Motoki and Wang, Bai Yang and Goodge, Berit H. and Harvey, Shannon P. and Lee, Kyuho and Li, Danfeng and Kourkoutis, Lena F. and Hwang, Harold Y.},
	month = nov,
	year = {2021},
	pages = {2104083},
}

@article{lee_linear--temperature_2023,
	title = {Linear-in-temperature resistivity for optimally superconducting ({Nd},{Sr}){NiO2}},
	volume = {619},
	issn = {0028-0836, 1476-4687},
	url = {https://www.nature.com/articles/s41586-023-06129-x},
	doi = {10.1038/s41586-023-06129-x},
	language = {en},
	number = {7969},
	urldate = {2023-10-23},
	journal = {Nature},
	author = {Lee, Kyuho and Wang, Bai Yang and Osada, Motoki and Goodge, Berit H. and Wang, Tiffany C. and Lee, Yonghun and Harvey, Shannon and Kim, Woo Jin and Yu, Yijun and Murthy, Chaitanya and Raghu, Srinivas and Kourkoutis, Lena F. and Hwang, Harold Y.},
	month = jul,
	year = {2023},
	pages = {288--292},
}

@article{lu_magnetic_2021,
	title = {Magnetic excitations in infinite-layer nickelates},
	volume = {373},
	issn = {0036-8075, 1095-9203},
	url = {https://www.science.org/doi/10.1126/science.abd7726},
	doi = {10.1126/science.abd7726},
	language = {en},
	number = {6551},
	urldate = {2025-01-27},
	journal = {Science},
	author = {Lu, H. and Rossi, M. and Nag, A. and Osada, M. and Li, D. F. and Lee, K. and Wang, B. Y. and Garcia-Fernandez, M. and Agrestini, S. and Shen, Z. X. and Been, E. M. and Moritz, B. and Devereaux, T. P. and Zaanen, J. and Hwang, H. Y. and Zhou, Ke-Jin and Lee, W. S.},
	month = jul,
	year = {2021},
	pages = {213--216},
}

@article{ortiz_magnetic_2022,
  title={Magnetic correlations in infinite-layer nickelates: An experimental and theoretical multimethod study},
  author={Ortiz, Roberto Antonio and Puphal, Pascal and Klett, Marcel and Hotz, Fabian and Kremer, Reinhard K and Trepka, Heiko and Hemmida, Mamoun and von Nidda, H-A Krug and Isobe, Masaaki and Khasanov, Rustem and others},
  journal={Physical Review Research},
  volume={4},
  number={2},
  pages={023093},
  year={2022},
  publisher={APS}
}

@article{kiaba_observation_2024,
	title = {Observation of {Mermin}-{Wagner} behavior in {LaFeO3}/{SrTiO3} superlattices},
	volume = {15},
	issn = {2041-1723},
	url = {https://www.nature.com/articles/s41467-024-49518-0},
	doi = {10.1038/s41467-024-49518-0},
	language = {en},
	number = {1},
	urldate = {2025-03-09},
	journal = {Nature Communications},
	author = {Kiaba, M. and Suter, A. and Salman, Z. and Prokscha, T. and Chen, B. and Koster, G. and Dubroka, A.},
	month = jun,
	year = {2024},
	pages = {5313},
}

@book{amato2024,
    author = {Amato, A. and Morenzoni, E.},
    title = {Introduction to Muon Spin Spectroscopy},
    publisher = {Springer Cham, Switzerland},
    year = {2024},
    doi = {10.1007/978-3-031-44959-8}
}

@article{simoes_muon_2020,
	title = {Muon implantation experiments in films: {Obtaining} depth-resolved information},
	volume = {91},
	issn = {0034-6748, 1089-7623},
	shorttitle = {Muon implantation experiments in films},
	url = {https://pubs.aip.org/rsi/article/91/2/023906/364854/Muon-implantation-experiments-in-films-Obtaining},
	doi = {10.1063/1.5126529},
	language = {en},
	number = {2},
	urldate = {2025-03-28},
	journal = {Review of Scientific Instruments},
	author = {Simões, A. F. A. and Alberto, H. V. and Vilão, R. C. and Gil, J. M. and Cunha, J. M. V. and Curado, M. A. and Salomé, P. M. P. and Prokscha, T. and Suter, A. and Salman, Z.},
	month = feb,
	year = {2020},
	pages = {023906},
}

@article{hayward_synthesis_2003,
	title = {Synthesis of the infinite layer {Ni}({I}) phase {NdNiO2}+x by low temperature reduction of {NdNiO3} with sodium hydride},
	volume = {5},
	copyright = {https://www.elsevier.com/tdm/userlicense/1.0/},
	issn = {12932558},
	url = {https://linkinghub.elsevier.com/retrieve/pii/S1293255803001110},
	doi = {10.1016/S1293-2558(03)00111-0},
	language = {en},
	number = {6},
	urldate = {2025-05-22},
	journal = {Solid State Sciences},
	author = {Hayward, M.A. and Rosseinsky, M.J.},
	month = jun,
	year = {2003},
	pages = {839--850},
}

@article{saykin_spin-glass_2025,
	title = {Spin-glass state in nickelate superconductors},
	volume = {10},
	issn = {2397-4648},
	url = {https://www.nature.com/articles/s41535-025-00813-z},
	doi = {10.1038/s41535-025-00813-z},
	language = {en},
	number = {1},
	urldate = {2025-12-09},
	journal = {npj Quantum Materials},
	author = {Saykin, David R. and Gonzalez, Martin and Fowlie, Jennifer and Kivelson, Steven A. and Hwang, Harold Y. and Kapitulnik, Aharon},
	month = aug,
	year = {2025},
	pages = {94},
}

@article{lin_universal_2022,
	title = {Universal spin-glass behaviour in bulk {LaNiO}$_{\textrm{2}}$ , {PrNiO}$_{\textrm{2}}$ and {NdNiO}$_{\textrm{2}}$},
	volume = {24},
	issn = {1367-2630},
	url = {https://iopscience.iop.org/article/10.1088/1367-2630/ac465e},
	doi = {10.1088/1367-2630/ac465e},
	language = {en},
	number = {1},
	urldate = {2025-12-11},
	journal = {New Journal of Physics},
	author = {Lin, Hai and Gawryluk, Dariusz Jakub and Klein, Yannick Maximilian and Huangfu, Shangxiong and Pomjakushina, Ekaterina and Von Rohr, Fabian and Schilling, Andreas},
	month = jan,
	year = {2022},
	pages = {013022},
}

@article{keimer_quantum_2015,
	title = {From quantum matter to high-temperature superconductivity in copper oxides},
	volume = {518},
	issn = {0028-0836, 1476-4687},
	url = {https://www.nature.com/articles/nature14165},
	doi = {10.1038/nature14165},
	language = {en},
	number = {7538},
	urldate = {2025-12-11},
	journal = {Nature},
	author = {Keimer, B. and Kivelson, S. A. and Norman, M. R. and Uchida, S. and Zaanen, J.},
	month = feb,
	year = {2015},
	pages = {179--186},
}

@article{scalapino_common_2012,
	title = {A common thread: {The} pairing interaction for unconventional superconductors},
	volume = {84},
	copyright = {http://link.aps.org/licenses/aps-default-license},
	issn = {0034-6861, 1539-0756},
	shorttitle = {A common thread},
	url = {https://link.aps.org/doi/10.1103/RevModPhys.84.1383},
	doi = {10.1103/RevModPhys.84.1383},
	language = {en},
	number = {4},
	urldate = {2025-12-11},
	journal = {Reviews of Modern Physics},
	author = {Scalapino, D. J.},
	month = oct,
	year = {2012},
	pages = {1383--1417},
}

@article{hepting_electronic_2020,
	 title={Electronic structure of the parent compound of superconducting infinite-layer nickelates},
  author={Hepting, Matthias and Li, Danfeng and Jia, CJ and Lu, Haiyu and Paris, E and Tseng, Y and Feng, X and Osada, M and Been, E and Hikita, Y and others},
  journal={Nature Materials},
  volume={19},
  number={4},
  pages={381--385},
  year={2020},
  publisher={Nature Publishing Group UK London}
}

@article{hepting2021soft,
  title={Soft X-ray spectroscopy of low-valence nickelates},
  author={Hepting, Matthias and Dean, Mark PM and Lee, Wei-Sheng},
  journal={Frontiers in Physics},
  volume={9},
  pages={808683},
  year={2021},
  publisher={Frontiers Media SA}
}

@inproceedings{plakida_spin-fluctuation_2001,
  title={Spin-fluctuation pairing in strongly correlated systems},
  author={Plakida, Nikolai M},
  booktitle={AIP Conference Proceedings},
  volume={580},
  number={1},
  pages={121--187},
  year={2001},
  organization={American Institute of Physics}
}

@article{dean_persistence_2013,
	title = {Persistence of magnetic excitations in {La2}−{xSrxCuO4} from the undoped insulator to the heavily overdoped non-superconducting metal},
	volume = {12},
	issn = {1476-1122, 1476-4660},
	url = {https://www.nature.com/articles/nmat3723},
	doi = {10.1038/nmat3723},
	language = {en},
	number = {11},
	urldate = {2025-12-12},
	journal = {Nature Materials},
	author = {Dean, M. P. M. and Dellea, G. and Springell, R. S. and Yakhou-Harris, F. and Kummer, K. and Brookes, N. B. and Liu, X. and Sun, Y-J. and Strle, J. and Schmitt, T. and Braicovich, L. and Ghiringhelli, G. and Božović, I. and Hill, J. P.},
	month = nov,
	year = {2013},
	pages = {1019--1023},
}

@article{goodge_doping_2021,
	title = {Doping evolution of the {Mott}–{Hubbard} landscape in infinite-layer nickelates},
	volume = {118},
	issn = {0027-8424, 1091-6490},
	url = {https://pnas.org/doi/full/10.1073/pnas.2007683118},
	doi = {10.1073/pnas.2007683118},
	language = {en},
	number = {2},
	urldate = {2025-12-12},
	journal = {Proceedings of the National Academy of Sciences},
	author = {Goodge, Berit H. and Li, Danfeng and Lee, Kyuho and Osada, Motoki and Wang, Bai Yang and Sawatzky, George A. and Hwang, Harold Y. and Kourkoutis, Lena F.},
	month = jan,
	year = {2021},
	pages = {e2007683118},
}

@article{bednorz1986possible,
  title={Possible high T c superconductivity in the Ba- La- Cu- O system},
  author={Bednorz, J George and M{\"u}ller, K Alex},
  journal={Zeitschrift f{\"u}r Physik B Condensed Matter},
  volume={64},
  number={2},
  pages={189--193},
  year={1986},
  publisher={Springer}
}

@article{li2019superconductivity,
  title={Superconductivity in an infinite-layer nickelate},
  author={Li, Danfeng and Lee, Kyuho and Wang, Bai Yang and Osada, Motoki and Crossley, Samuel and Lee, Hye Ryoung and Cui, Yi and Hikita, Yasuyuki and Hwang, Harold Y},
  journal={Nature},
  volume={572},
  number={7771},
  pages={624--627},
  year={2019},
  publisher={Nature Publishing Group UK London}
}

@article{le2011intense,
  title={Intense paramagnon excitations in a large family of high-temperature superconductors},
  author={Le Tacon, Mathieu and Ghiringhelli, G and Chaloupka, J and Sala, M Moretti and Hinkov, V and Haverkort, MW and Minola, Matteo and Bakr, M and Zhou, KJ and Blanco-Canosa, S and others},
  journal={Nature Physics},
  volume={7},
  number={9},
  pages={725--730},
  year={2011},
  publisher={Nature Publishing Group UK London}
}

@article{julien2003magnetic,
  title={Magnetic order and superconductivity in La2- xSrxCuO4: a review},
  author={Julien, M-H},
  journal={Physica B: Condensed Matter},
  volume={329},
  pages={693--696},
  year={2003},
  publisher={Elsevier}
}

@article{hasselmann2004spin,
  title={Spin-glass phase of cuprates},
  author={Hasselmann, N and Neto, AH Castro and Smith, C Morais},
  journal={Physical Review B},
  volume={69},
  number={1},
  pages={014424},
  year={2004},
  publisher={APS}
}

@article{aharony1988magnetic,
    title={Magnetic phase diagram and magnetic pairing in doped La 2 CuO 4},
  author={Aharony, Amnon and Birgeneau, RJ and Coniglio, A and Kastner, MA and Stanley, HE},
  journal={Physical Review Letters},
  volume={60},
  number={13},
  pages={1330},
  year={1988},
  publisher={APS}
}

@article{lee2025effects,
  title={Effects of stoichiometry and epitaxial strain on the stabilization of infinite-layer nickelates},
  author={Lee, Kyuho and Goodge, Berit H and Lee, Yonghun and Kim, Woo Jin and Osada, Motoki and Wang, Bai Yang and Wang, Tiffany C and Hwang, Harold Y},
  journal={APL Materials},
  volume={13},
  number={10},
  year={2025},
  publisher={AIP Publishing}
}

@article{osada2023improvement,
  title={Improvement of superconducting properties in La 1- x Sr x Ni O 2 thin films by tuning topochemical reduction temperature},
  author={Osada, Motoki and Fujiwara, Kohei and Nojima, Tsutomu and Tsukazaki, Atsushi},
  journal={Physical Review Materials},
  volume={7},
  number={5},
  pages={L051801},
  year={2023},
  publisher={APS}
}

@article{gonzalez2024absence,
  title={Absence of hydrogen insertion into highly crystalline superconducting infinite layer nickelates},
  author={Gonzalez, Martin and Ievlev, A and Lee, K and Kim, W and Yu, Y and Fowlie, J and Hwang, HY},
  journal={Physical Review Materials},
  volume={8},
  number={8},
  pages={084804},
  year={2024},
  publisher={APS}
}

@article{lee2020aspects,
  title={Aspects of the synthesis of thin film superconducting infinite-layer nickelates},
  author={Lee, Kyuho and Goodge, Berit H and Li, Danfeng and Osada, Motoki and Wang, Bai Yang and Cui, Yi and Kourkoutis, Lena F and Hwang, Harold Y},
  journal={APL Materials},
  volume={8},
  number={4},
  year={2020},
  publisher={AIP Publishing}
}

@article{morenzoni2002implantation,
  title={Implantation studies of keV positive muons in thin metallic layers},
  author={Morenzoni, E and Gl{\"u}ckler, H and Prokscha, T and Khasanov, R and Luetkens, H and Birke, M and Forgan, EM and Niedermayer, Ch and Pleines, M},
  journal={Nuclear Instruments and Methods in Physics Research Section B: Beam Interactions with Materials and Atoms},
  volume={192},
  number={3},
  pages={254--266},
  year={2002},
  publisher={Elsevier}
}

@book{eckstein2013computer,
  title={Computer simulation of ion-solid interactions},
  author={Eckstein, Wolfgang},
  volume={10},
  year={2013},
  publisher={Springer Science \& Business Media}
}

@article{salman2014direct,
  title={Direct spectroscopic observation of a shallow hydrogenlike donor state in insulating SrTiO 3},
  author={Salman, Z and Prokscha, T and Amato, A and Morenzoni, E and Scheuermann, R and Sedlak, K and Suter, A},
  journal={Physical Review Letters},
  volume={113},
  number={15},
  pages={156801},
  year={2014},
  publisher={APS}
}

@inproceedings{suter2023low,
  title={Low Energy Measurements in Low-Energy $\mu$SR},
  author={Suter, Andreas and Martins, Maria Mendes and Ni, Xiaojie and Prokscha, Thomas and Salman, Zaher},
  booktitle={Journal of Physics: Conference Series},
  volume={2462},
  number={1},
  pages={012011},
  year={2023},
  organization={IOP Publishing}
}

@book{blundell2021muon,
  title={Muon spectroscopy: an introduction},
  author={Blundell, Stephen and De Renzi, Roberto and Lancaster, Tom and Pratt, Francis L},
  year={2021},
  publisher={Oxford University Press}
}

@article{hayano1979zero,
  title={Zero-and low-field spin relaxation studied by positive muons},
  author={Hayano, RS and Uemura, YJ and Imazato, J and Nishida, N and Yamazaki, T and Kubo, R},
  journal={Physical Review B},
  volume={20},
  number={3},
  pages={850},
  year={1979},
  publisher={APS}
}

@article{borsa1995staggered,
  title={Staggered magnetization in La 2- x Sr x CuO 4 from La 139 NQR and $\mu$SR: Effects of Sr doping in the range 0< x< 0.02},
  author={Borsa, Ferdinando and Carretta, Pietro and Cho, JH and Chou, FC and Hu, Q and Johnston, DC and Lascialfari, A and Torgeson, DR and Gooding, RJ and Salem, NM and others},
  journal={Physical Review B},
  volume={52},
  number={10},
  pages={7334},
  year={1995},
  publisher={APS}
}

@book{le2011muon,
  title={Muon spin rotation, relaxation, and resonance: applications to condensed matter},
  author={Le Yaouanc, Alain and De Reotier, Pierre Dalmas},
  number={147},
  year={2011},
  publisher={OUP Oxford}
}

@article{suter2012musrfit,
  title={Musrfit: a free platform-independent framework for $\mu$SR data analysis},
  author={Suter, A and Wojek, BM},
  journal={Physics Procedia},
  volume={30},
  pages={69--73},
  year={2012},
  publisher={Elsevier}
}

@article{de1992quantum,
  title={Quantum calculation of the muon depolarization function: effect of spin dynamics in nuclear dipole systems},
  author={de R{\'e}otier, P Dalmas and Yaouanc, A},
  journal={Journal of Physics: Condensed Matter},
  volume={4},
  number={18},
  pages={4533},
  year={1992},
  publisher={IOP Publishing}
}

@article{keren1994generalization,
  title={Generalization of the Abragam relaxation function to a longitudinal field},
  author={Keren, Amit},
  journal={Physical Review B},
  volume={50},
  number={14},
  pages={10039},
  year={1994},
  publisher={APS}
}

@article{campbell1994dynamics,
  title={Dynamics in canonical spin glasses observed by muon spin depolarization},
  author={Campbell, IA and Amato, A and Gygax, FN and Herlach, D and Schenck, A and Cywinski, R and Kilcoyne, SH},
  journal={Physical Review Letters},
  volume={72},
  number={8},
  pages={1291},
  year={1994},
  publisher={APS}
}

@article{lichtensteiger2018interactivexrdfit,
  title={InteractiveXRDFit: a new tool to simulate and fit X-ray diffractograms of oxide thin films and heterostructures},
  author={Lichtensteiger, C{\'e}line},
  journal={Applied Crystallography},
  volume={51},
  number={6},
  pages={1745--1751},
  year={2018},
  publisher={International Union of Crystallography}
}

@article{ito2023understanding,
  title={Understanding muon diffusion in perovskite oxides below room temperature based on harmonic transition state theory},
  author={Ito, TU and Higemoto, W and Shimomura, K},
  journal={Physical Review B},
  volume={108},
  number={22},
  pages={224301},
  year={2023},
  publisher={APS}
}

@article{holzschuh1983muon,
  title={Muon-spin-rotation experiments in orthoferrites},
  author={Holzschuh, E and Denison, AB and K{\"u}ndig, W and Meier, PF and Patterson, BD},
  journal={Physical Review B},
  volume={27},
  number={9},
  pages={5294},
  year={1983},
  publisher={APS}
}

@article{hempelmann1998muon,
  title={Muon diffusion and trapping in proton conducting oxides},
  author={Hempelmann, Rolf and Soetratmo, M and Hartmann, Olivier and W{\"a}ppling, Roger},
  journal={Solid State Ionics},
  volume={107},
  number={3-4},
  pages={269--280},
  year={1998},
  publisher={Elsevier}
}

@article{ito2017excited,
  title={Excited configurations of hydrogen in the BaTiO 3- x H x perovskite lattice associated with hydrogen exchange and transport},
  author={Ito, TU and Koda, A and Shimomura, K and Higemoto, W and Matsuzaki, T and Kobayashi, Y and Kageyama, H},
  journal={Physical Review B},
  volume={95},
  number={2},
  pages={020301},
  year={2017},
  publisher={APS}
}

@article{blundell1999spin,
  title={Spin-polarized muons in condensed matter physics},
  author={Blundell, SJ},
  journal={Contemporary Physics},
  volume={40},
  number={3},
  pages={175--192},
  year={1999},
  publisher={Taylor \& Francis}
}

@article{sun2025electronic,
  title={Electronic structure of superconducting infinite-layer lanthanum nickelates},
  author={Sun, Wenjie and Jiang, Zhicheng and Xia, Chengliang and Hao, Bo and Yan, Shengjun and Wang, Maosen and Li, Yueying and Liu, Hongquan and Ding, Jianyang and Liu, Jiayu and others},
  journal={Science Advances},
  volume={11},
  number={4},
  pages={eadr5116},
  year={2025},
  publisher={American Association for the Advancement of Science}
}

@article{wang2025molecular,
  title={Molecular H2 as the Reducing Agent in Low-Temperature Oxide Reduction Using Calcium Hydride},
  author={Wang, Jiayue and Yu, Yijun and Abdelkawy, Ahmed and Li, Jiarui and Li, Jinlei and Yang, Jing and Ko, Eun Kyo and Lee, Yonghun and Thampy, Vivek and Cui, Yi and others},
  journal={Journal of the American Chemical Society},
  volume={147},
  number={4},
  pages={3032--3038},
  year={2025},
  publisher={ACS Publications}
}

@article{yamamoto2013hydride,
  title={Hydride reductions of transition metal oxides},
  author={Yamamoto, Takafumi and Kageyama, Hiroshi},
  journal={Chemistry letters},
  volume={42},
  number={9},
  pages={946--953},
  year={2013},
  publisher={Oxford University Press}
}

@article{murnick1976muon,
  title={Muon-spin depolarization in spin-glasses},
  author={Murnick, DE and Fiory, AT and Kossler, WJ},
  journal={Physical Review Letters},
  volume={36},
  number={2},
  pages={100},
  year={1976},
  publisher={APS}
}

@article{emmerich1981mu+,
  title={$\mu$+ investigations of spin freezing processes in CuMn},
  author={Emmerich, K and Schwink, Ch},
  journal={Hyperfine Interactions},
  volume={8},
  number={4},
  pages={767--770},
  year={1981},
  publisher={Springer}
}

@article{brown1981anomalous,
  title={Anomalous paramagnetic-state $\mu$SR in spin-glass AgMn},
  author={Brown, JA and Heffner, RH and Kitchens, TA and Leon, M and Olsen, CE and Schillaci, ME and Dodds, SA and MacLaughlin, DE},
  journal={Journal of Applied Physics},
  volume={52},
  number={3},
  pages={1766--1768},
  year={1981},
  publisher={American Institute of Physics}
}

@article{bernhard2012muon,
  title={Muon spin rotation study of magnetism and superconductivity in Ba (Fe 1- x Co x) 2 As 2 single crystals},
  author={Bernhard, Christian and Wang, Chen Nan and Nuccio, Laura and Schulz, Leander and Zaharko, O and Larsen, Jacob and Aristizabal, C and Willis, M and Drew, Alan J and Varma, Ghanshyam Das and others},
  journal={Physical Review B Condensed Matter and Materials Physics},
  volume={86},
  number={18},
  pages={184509},
  year={2012},
  publisher={APS}
}

@article{takeshita2009competition,
  title={Competition/coexistence of magnetism and superconductivity in iron pnictides probed by muon spin rotation},
  author={Takeshita, Soshi and Kadono, Ryosuke},
  journal={New Journal of Physics},
  volume={11},
  number={3},
  pages={035006},
  year={2009}
}

@article{vaknin1987antiferromagnetism,
  title={Antiferromagnetism in La 2 CuO 4- y},
  author={Vaknin, D and Sinha, SK and Moncton, DE and Johnston, DC and Newsam, JM and Safinya, CR and King Jr, HE},
  journal={Physical Review Letters},
  volume={58},
  number={26},
  pages={2802},
  year={1987},
  publisher={APS}
}

@article{tsuei2000pairing,
  title={Pairing symmetry in cuprate superconductors},
  author={Tsuei, CC and Kirtley, JR},
  journal={Reviews of Modern Physics},
  volume={72},
  number={4},
  pages={969},
  year={2000},
  publisher={APS}
}

@article{harvey2025evidence,
  title={Evidence for nodal superconductivity in infinite-layer nickelates},
  author={Harvey, Shannon P and Wang, Bai Yang and Fowlie, Jennifer and Osada, Motoki and Lee, Kyuho and Lee, Yonghun and Li, Danfeng and Hwang, Harold Y},
  journal={Proceedings of the National Academy of Sciences},
  volume={122},
  number={48},
  pages={e2427243122},
  year={2025},
  publisher={National Academy of Sciences}
}

@article{tallon1997muon,
  title={Muon spin relaxation studies of superconducting cuprates},
  author={Tallon, Jeffery L and Bernhard, Christian and Niedermayer, Christof},
  journal={Superconductor Science and Technology},
  volume={10},
  number={7A},
  pages={A38--A51},
  year={1997}
}

@article{stilp2013magnetic,
  title={Magnetic phase diagram of low-doped La2-xSrxCuO4 thin films studied by low-energy muon-spin rotation},
  author={Stilp, E and Suter, A and Prokscha, T and Morenzoni, E and Keller, H and Wojek, BM and Luetkens, H and Gozar, A and Logvenov, G and Bo{\v{z}}ovi{\'c}, I},
  journal={Physical Review B},
  volume={88},
  year={2013}
}

@article{binder1986spin,
  title={Spin glasses: Experimental facts, theoretical concepts, and open questions},
  author={Binder, Kurt and Young, A Peter},
  journal={Reviews of Modern physics},
  volume={58},
  number={4},
  pages={801},
  year={1986},
  publisher={APS}
}

@article{ruderman1954indirect,
  title={Indirect exchange coupling of nuclear magnetic moments by conduction electrons},
  author={Ruderman, Melvin A and Kittel, Charles},
  journal={Physical Review},
  volume={96},
  number={1},
  pages={99},
  year={1954},
  publisher={APS}
}

@article{kastner1998magnetic,
  title={Magnetic, transport, and optical properties of monolayer copper oxides},
  author={Kastner, MA and Birgeneau, RJ and Shirane, G and Endoh, Y},
  journal={Reviews of Modern Physics},
  volume={70},
  number={3},
  pages={897},
  year={1998},
  publisher={APS}
}

@article{fowlie2023metal,
  title={Metal--insulator transition in composition-tuned nickel oxide films},
  author={Fowlie, Jennifer and Georgescu, Alexandru B and Suter, Andreas and Mundet, Bernat and Toulouse, Constance and Jaouen, Nicolas and Viret, Michel and Dom{\'\i}nguez, Claribel and Gibert, Marta and Salman, Zaher and others},
  journal={Journal of Physics: Condensed Matter},
  volume={35},
  number={30},
  pages={304001},
  year={2023},
  publisher={IOP Publishing}
}

@article{zhou2025origin,
  title={Origin of local magnetic exchange interaction in infiite-layer nickelates},
  author={Zhou, Yanbing and Zhao, Dan and Zeng, Boyun and Xia, Chengliang and Wang, Yu and Chen, Hanghui and Wu, Tao and Chen, Xianhui},
  journal={arXiv preprint arXiv:2505.09476},
  year={2025}
}

@article{pal2018quasistatic,
  title={Quasistatic internal magnetic field detected in the pseudogap phase of Bi 2+ x Sr 2- x CaCu 2 O 8+ $\delta$ by muon spin relaxation},
  author={Pal, Anand and Dunsiger, SR and Akintola, Kolawole and Fang, ACY and Elhosary, Abdo and Ishikado, Motoyuki and Eisaki, Hiroshi and Sonier, JE},
  journal={Physical Review B},
  volume={97},
  number={6},
  pages={060502},
  year={2018},
  publisher={APS}
}

@article{sonier2002correlations,
  title={Correlations between charge ordering and local magnetic fields in overdoped YBa 2 Cu 3 O 6+ x},
  author={Sonier, JE and Brewer, JH and Kiefl, RF and Heffner, RH and Poon, KF and Stubbs, SL and Morris, GD and Miller, RI and Hardy, WN and Liang, R and others},
  journal={Physical Review B},
  volume={66},
  number={13},
  pages={134501},
  year={2002},
  publisher={APS}
}

@article{ranna2025disorder,
  title={Disorder-induced suppression of superconductivity in infinite-layer nickelates},
  author={Ranna, Abhishek and Grasset, Romain and Gonzalez, Martin and Lee, Kyuho and Wang, Bai Yang and Abarca Morales, Edgar and Theuss, Florian and Filipiak, Zuzanna H and Moravec, Michal and Konczykowski, Marcin and others},
  journal={Physical Review Letters},
  volume={135},
  number={12},
  pages={126501},
  year={2025},
  publisher={APS}
}

@article{ding2024cuprate,
  title={Cuprate-like electronic structures in infinite-layer nickelates with substantial hole dopings},
  author={Ding, Xiang and Fan, Yu and Wang, Xiaoxiao and Li, Chihao and An, Zhitong and Ye, Jiahao and Tang, Shenglin and Lei, Minyinan and Sun, Xingtian and Guo, Nan and others},
  journal={National Science Review},
  volume={11},
  number={8},
  pages={nwae194},
  year={2024},
  publisher={Oxford University Press}
}

@article{Osada_Hall_LSNO_2025,
 title={Systematic evolution of superconducting pairing strength and Seebeck coefficients in correlated infinite-layer La1--x Sr x NiO2},
  author={Osada, Motoki and Imajo, Shusaku and Seki, Yuji and Ishida, Kousuke and Nojima, Tsutomu and Fujiwara, Kohei and Kindo, Koichi and Nomura, Yusuke and Tsukazaki, Atsushi},
  journal={Science Advances},
  volume={11},
  number={40},
  pages={eadv6488},
  year={2025},
  publisher={American Association for the Advancement of Science}
}

@article{ryan2004zero,
  title={Zero-field muon spin relaxation studies of frustrated magnets: physics and analysis issues},
  author={Ryan, DH and Van Lierop, J and Cadogan, JM},
  journal={Journal of Physics: Condensed Matter},
  volume={16},
  number={40},
  pages={S4619--S4638},
  year={2004}
}

@article{kim2026optical,
  title={Optical conductivity signatures of strong correlations and multiband superconductivity in infinite-layer nickelates},
  author={Kim, Woo Jin and Lee, Kyuho and Ko, Eun Kyo and Son, Jaeseok and Lee, Yonghun and Yu, Yijun and Moon, Soon Jae and Noh, Tae Won and Hwang, Harold Y},
  journal={arXiv preprint arXiv:2602.09567},
  year={2026}
}

@article{rossi2021orbital,
  title={Orbital and spin character of doped carriers in infinite-layer nickelates},
  author={Rossi, Matteo and Lu, Haiyu and Nag, Abhishek and Li, Danfeng and Osada, Motoki and Lee, Kyuho and Wang, Bai Yang and Agrestini, Stefano and Garcia-Fernandez, M and Kas, JJ and others},
  journal={Physical Review B},
  volume={104},
  number={22},
  pages={L220505},
  year={2021},
  publisher={APS}
}

@article{pan2026superconducting,
  title={Superconducting phase diagram of multilayer square-planar nickelates},
  author={Pan, Grace A and Segedin, Dan Ferenc and TenHuisen, Sophia FR and Bhatt, Lopa and LaBollita, Harrison and Jiang, Abigail Y and Song, Qi and Turkiewicz, Ari B and Baykusheva, Denitsa R and Nag, Abhishek and others},
  journal={Science},
  volume={392},
  number={6805},
  pages={1390--1395},
  year={2026},
  publisher={American Association for the Advancement of Science}
}

@article{lechermann2020multiorbital,
  title={Multiorbital processes rule the Nd 1-x Sr x NiO 2 normal state},
  author={Lechermann, Frank},
  journal={Physical Review X},
  volume={10},
  number={4},
  pages={041002},
  year={2020},
  publisher={APS}
}

@article{lechermann2020late,
  title={Late transition metal oxides with infinite-layer structure: Nickelates versus cuprates},
  author={Lechermann, Frank},
  journal={Physical Review B},
  volume={101},
  number={8},
  pages={081110},
  year={2020},
  publisher={APS}
}

@article{sonier2010high,
  title={High-field $\mu$SR studies of superconducting and magnetic correlations in cuprates above T c},
  author={Sonier, JE},
  journal={Journal of Physics: Condensed Matter},
  volume={22},
  number={20},
  pages={203202},
  year={2010}
}

@article{prokscha2008new,
  title={The new $\mu$E4 beam at PSI: A hybrid-type large acceptance channel for the generation of a high intensity surface-muon beam},
  author={Prokscha, T and Morenzoni, E and Deiters, K and Foroughi, F and George, D and Kobler, R and Suter, A and Vrankovic, V},
  journal={Nuclear Instruments and Methods in Physics Research Section A: Accelerators, Spectrometers, Detectors and Associated Equipment},
  volume={595},
  number={2},
  pages={317--331},
  year={2008},
  publisher={Elsevier}
}

@article{zhang1988effective,
  title={Effective Hamiltonian for the superconducting Cu oxides},
  author={Zhang, FC and Rice, TM},
  journal={Physical Review B},
  volume={37},
  number={7},
  pages={3759},
  year={1988},
  publisher={APS}
}

@article{cannella1972magnetic,
  title={Magnetic ordering in gold-iron alloys},
  author={Cannella, Vincent and Mydosh, John A},
  journal={Physical Review B},
  volume={6},
  number={11},
  pages={4220},
  year={1972},
  publisher={APS}
}

@article{mulder1981susceptibility,
  title={Susceptibility of the Cu Mn spin-glass: Frequency and field dependences},
  author={Mulder, CAM and Van Duyneveldt, AJ and Mydosh, JA},
  journal={Physical Review B},
  volume={23},
  number={3},
  pages={1384},
  year={1981},
  publisher={APS}
}

@article{tang2026ni,
  title={Ni-O hybridization-driven electronic reconstruction across the superconducting dome in an infinite-layer nickelate},
  author={Tang, Chi Sin and Zeng, Shengwei and Gao, Xing and Luo, Zhaoyang and Liu, Xiongfang and Lim, Zhi Shiuh and Prakash, Saurav and Yang, Ping and Diao, Caozheng and Yin, Xinmao and others},
  journal={arXiv preprint arXiv:2605.30752},
  year={2026}
}

\clearpage
\onecolumngrid
% Reset figure numbering with "S" prefix
\setcounter{figure}{0}
\renewcommand{\thefigure}{S\arabic{figure}}
\renewcommand{\theHfigure}{S\arabic{figure}}  % for hyperref links

\section*{Supplementary Information}
\section{A. Sample preparation and characterization}\label{app:samples}
The perovskite phase nickelates, (La,Sr)NiO$_3$, were grown using pulsed laser deposition (PLD) on single-crystal SrTiO$_3$ (001) substrates. The
substrates were cleaned with acetone/isopropyl alcohol ultra–sonication, and annealed at 900$^{\circ}$ C for 30 minutes under an oxygen partial pressure of 2 $\times$ 10$^{-6}$ Torr. The nickelate film and SrTiO$_3$ capping layer were deposited at 570$^{\circ}$ C under an oxygen partial pressure of 150 mTorr from  polycrystalline pressed powder targets. The excimer laser output was tuned to a fluence of 1.8 J cm$^{-2}$ and operated at a pulse rate of 4 Hz. The perovskite samples were grown on 10 mm x 10 mm sized substrates which were subsequently cut into smaller pieces to form a sample mosaic. The growth conditions for all samples in the doping series are equivalent.

\begin{figure}[htbp]
   % \captionsetup{width=\textwidth}
    \centering
    \includegraphics[width=0.77\textwidth]{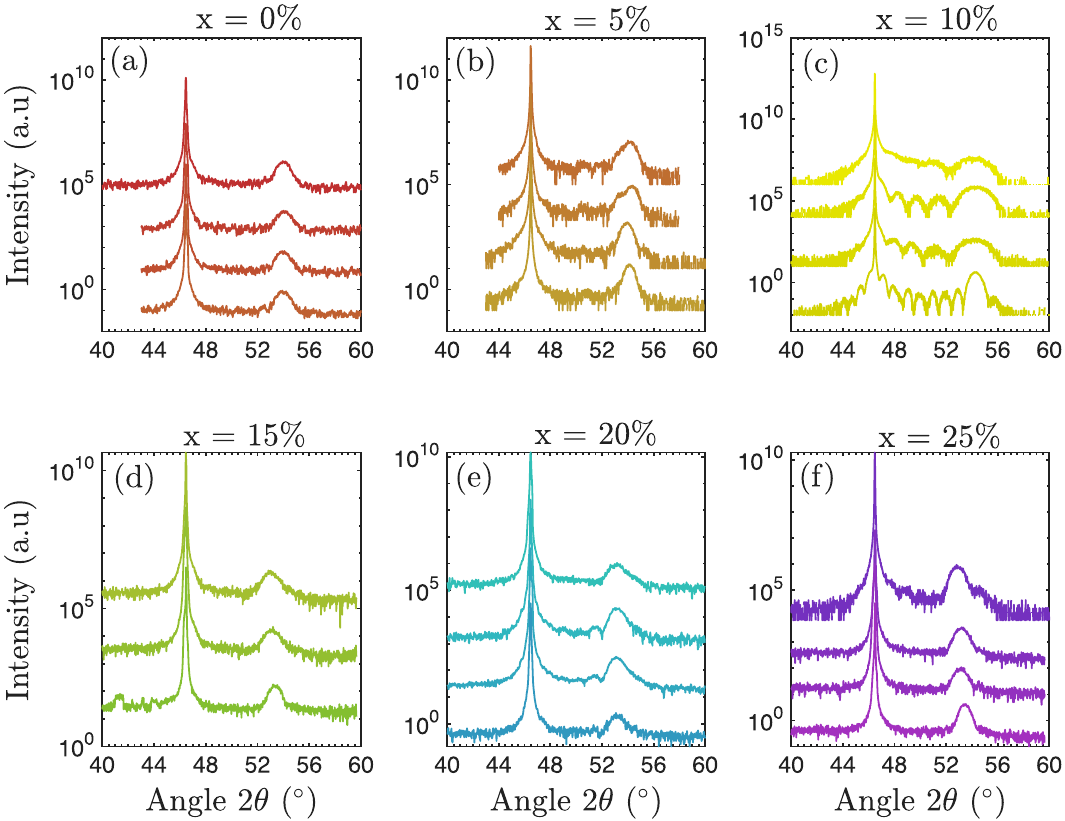}
    \caption{$2\theta$--$\omega$ symmetric scans of (La,Sr)NiO$_2$ thin films grown on SrTiO$_3$ (001) substrates. The XRD diffraction patterns are representative of hole-doped nickelate films with Sr doping levels (a) $x=0\%$, (b) $x=5 \%$, (c) $x=10\%$, (d) $x=15\%$, (e) $x=20\%$, and (f) $x=25\%$.}
    \label{fig:Supp_XRD}
\end{figure}

The perovskite precursor samples were then transformed into their infinite layer phase via apical oxygen de-intercalation induced by a topochemical reaction. The samples were wrapped in aluminum foil and reacted with 0.15 g CaH$_2$ powder within a vacuum sealed Pyrex glass tube (< 5 mTorr). The glass tube was then heated at 320$^{\circ}$ C for approximately 3 hours, and the infinite layer phase was verified by X-ray diffraction. Figure.~\ref{fig:Supp_XRD} shows the 2$\theta$-$\omega$ diffraction pattern for the samples used in the current study. While there is some variability in the thicknesses of the thin films, they all complete a full transition from the perovskite to the infinite layer phase. The reduced phase nickelates all have an extracted thickness range between 8 to 14 nm, as summarized in Table~\ref{tab:sample_summary}.

\begin{figure} [htbp]
   % \captionsetup{width=\textwidth}
    \centering
    \includegraphics[width=1\textwidth]{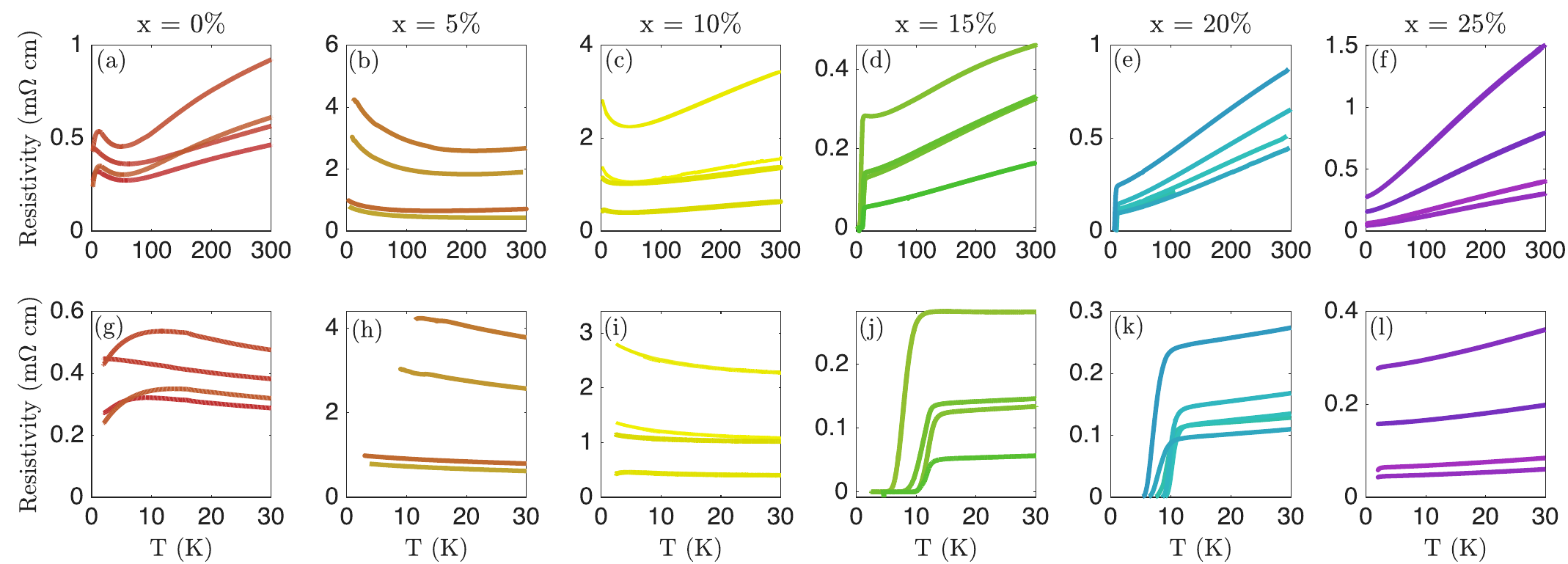}
    \caption{
    Temperature-dependent resistivity curves for the hole-doping series of the infinite layer nickelate \LSNO with (a, g) x = 0\%, (b, h) x = 5\%, (c, i) x = 10\%, (d, j) x =  15\%, (e, k) x = 20\%, and (f, l) x = 25\%. The top panel shows the full resistivity curves up to room temperature, whereas the bottom panel focuses on the low temperature regime up until 30 K. Data for the x = 0\% sample and x = 20\% doping levels are reproduced from our previous work ref~\cite{fowlie_intrinsic_2022}. }
    \label{fig:Supp_RT}
\end{figure}

Transport measurements on the samples used for this doping-dependent study were carried out using the four-probe geometry through Al-bonded wire contacts. Figure~\ref{fig:Supp_RT} shows the temperature-dependent resistivity, $\rho(T)$, for all samples used for each mosaic in the doping series. The resistivity curves for each doping level is consistent with the behavior reported in previous literature. The undoped nickelate shows an insulating upturn above ~50 K, with an onset of a superconducting downturn below 10 K. Within the underdoped regime (x = 5\%, 10\%) the room temperature resistivity shows metallic behavior followed by an insulating upturn at a temperature that is dependent on the doping level. Within the superconducting dome (x = 15\%, 20\%), samples exhibit superconductivity with T$_{c,0}$ dependent on the doping level. In the over-doped regime (x = 25\%),  we see metallicity down to the lowest temperatures.

\begin{figure}[h]
    \centering
	\includegraphics[width=\textwidth]{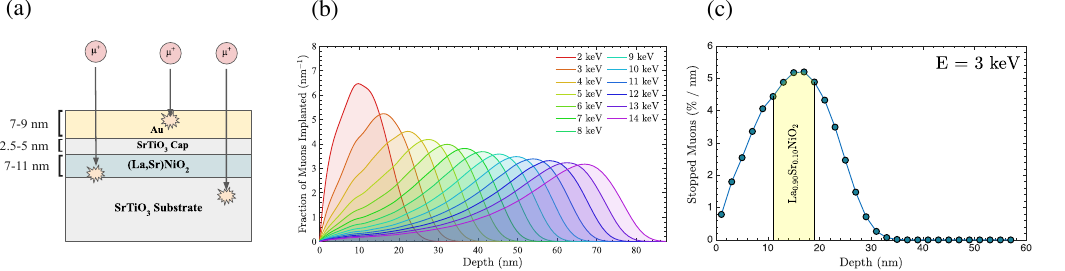}
	\caption{
        (a) The heterostructures consist of nickelate films grown on (001)-oriented SrTiO$_3$ substrates, capped with a protective SrTiO$_3$ capping layer, and deposited with gold via e-beam evaporation. (b) Monte Carlo simulations using TRIM.SP, showing the muon implantation profile for a sample heterostructure over a range of incident muon energies between 2 keV and 14 keV.  (c) Projected depth-dependent probability of muon implantation from the optimized sample heterostructure. 
    }
\label{fig:Supp_hetero}
\end{figure}

Our samples are designed to maximize muon implantation into the nickelate layer of the thin film heterostructure. Figure~\ref{fig:Supp_hetero}(a) shows a typical heterostructure consisting of the SrTiO$_3$ substrate, nickelate layer, SrTiO$_3$ cap, and an evaporated gold layer. For any given experiment, muons will implant in every layer of the heterostructure, however we can tune the energy of the incident muon beam to maximize the probability that a given muon will implant into the nickelate layer. Figure~\ref{fig:Supp_hetero}(b) shows the muon implantation profiles over a range of incident energies between 2 keV and 14 keV, which were estimated using Monte Carlo simulations \cite{eckstein2013computer}. We typically avoid muon beam energies below 2 keV because a substantial fraction of muons are lost to backscattering and reflection. Simultaneously we avoid muon energies above 4 keV because at higher energies, muons penetrate deeper into heterostructure and reduce implantation in the nickelate layer. Hence all experiments use a 3 keV muon incident beam, and the \STO capping layer and deposited gold are added to ensure implantation is maximized between 10 nm to 20 nm of the sample heterostructure. Figure~\ref{fig:Supp_hetero}(c) shows the depth-dependent probability of muon implantation, whose peak coincides with the nickelate layer. Hence, we maximize the muon signal arising from the nickelate layer.

\begin{table}[htbp]
\centering
\caption{Summary of average nickelate film parameters as a function of doping.}
\label{tab:nickelate_params}
\begin{tabular}{|c|c|c|c|c|c|}
\hline
Doping & Au Cap Thickness (nm) & STO Cap Thickness (nm) &Nickelate Thickness (nm) & $c$-axis lattice constant (\AA) & $T_{c,0}$ (K) \\
\hline
0\%  & - & 8.9 & 9.1 & 3.391 & - \\
\hline
5\%  & 7 & 2.5  &  11.4  &  3.390     & - \\
\hline
10\%  & 9 & 2.5 & 6.6  
&   3.379    & - \\
\hline
15\% & 7 & 4.3  &   8.2   &   3.438    & 6.6 \\
\hline
20\%  & - & 13.6 & 7.7 & 3.443 & 7.4 \\
\hline
25\%  & 9 & 2.5 &  7.3   &   3.447  & - \\
\hline
\end{tabular}
\label{tab:sample_summary}
\end{table}

Table~\ref{tab:sample_summary} summarizes the mean thicknesses, $c$-axis lattice constant, and critical temperatures for each doping level. The reported value for each doping level is averaged between the different pieces used to make each sample mosaic of at least 1 cm$^2$ surface area. The $c$-axis lattice parameters and thicknesses are obtained from fitting of the XRD 2$\theta$-$\omega$ diffraction patterns using a MATLAB routine \cite{lichtensteiger2018interactivexrdfit}. The transition temperature, T$_{c,0}$, is defined as the temperature where the resistance reaches the noise floor. Note that data for the x = 0\% sample and x = 20\% doping levels are reproduced from our previous work ref~\cite{fowlie_intrinsic_2022}. Note that our calculated $F_{\rm M}$ subtracts asymmetry contributions from the SrTiO$_3$ substrate~\cite{salman2014direct} and from the background asymmetry of the sample plate and differs from sample to sample due to the variation in the thicknesses of the nickelate and capping layers.

\begin{figure}[htbp]
    \centering
	\includegraphics[width=0.72\textwidth]{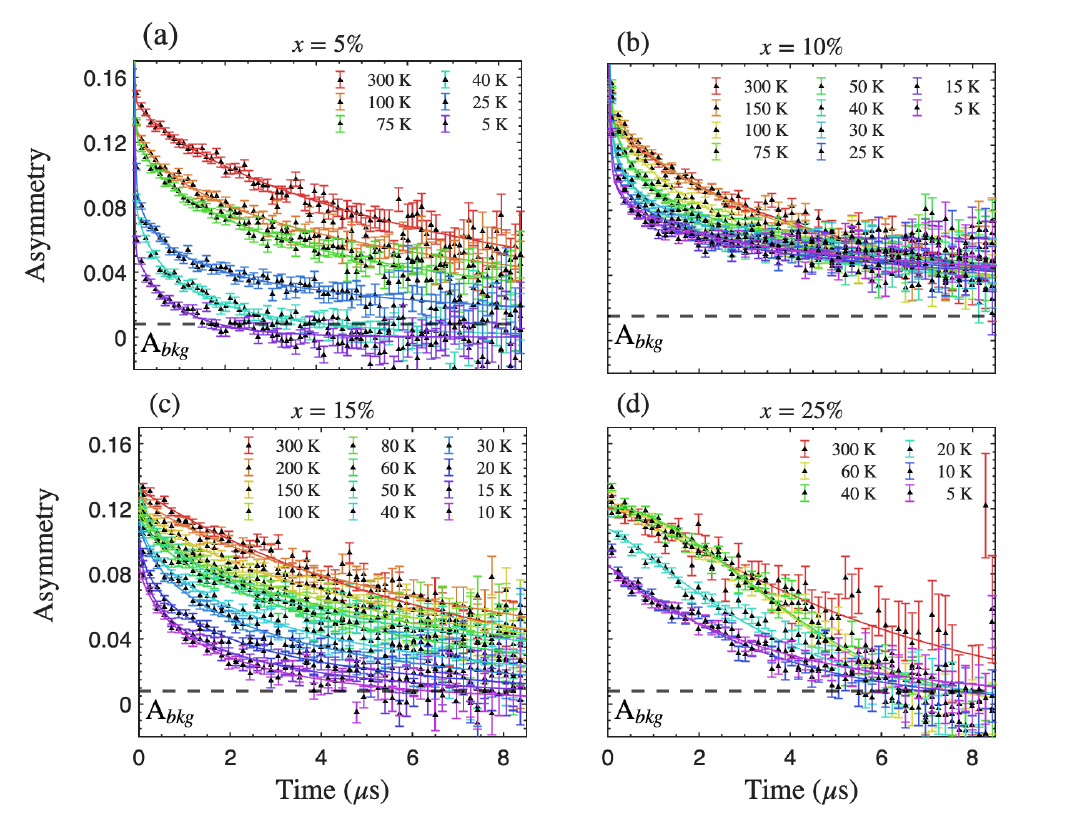}
	\caption{The full temperature-dependent ZF spectra at doping levels (a) x = 5\%, (b) x = 10\%, (c) x = 15\%, and (d) x = 25\%. As the temperature decreases, the shape of the ZF spectra continuously dampens, reflecting the changing magnetic ground state of the system.}
\label{fig:Supp_ZF}
\end{figure}

\section{B. Temperature Dependent ZF Spectra}
Figure~\ref{fig:Supp_ZF} shows the full temperature evolution of the ZF spectra for the x = 5\%, 10\%, 15\%, and 25\% doped samples. The temperature-dependent ZF spectra for each doping level shows a change in asymmetry shape towards a more damped exponential at lower temperatures. The increasingly stretched shape of the ZF spectra at low temperatures is indicative of the emergence of static magnetism from the freezing of electronic moments and the transition into a spin glass. This general trend is seen for all doping levels and is also consistent with ZF measurements on bulk powder \LNO samples as shown in ref.~\cite{ortiz_magnetic_2022}.

We bring particular attention to the room-temperature ZF spectra of the x = 25\% doped sample that shows evidence of muon diffusion. In oxide systems, implanted muons are assumed to be stationary objects under cryogenic temperatures. This is
because the muon is believed to be tightly bound to an oxygen anion in the host lattice~\cite{ito2023understanding}. However, at elevated temperatures, implanted muons may become mobile and will move through the sample's crystal lattice. In this scenario, the implanted muon will experience spatially differing local fields which will affect its depolarization. This phenomenon is known as motional narrowing, where the muon response is averaged over spatially varying fields and its spin relaxation is reduced~\cite{blundell1999spin}. Muon diffusion has been observed in orthoferrites~\cite{holzschuh1983muon}, Sc-doped SrZrO$_3$~\cite{hempelmann1998muon}, cuprates~\cite{sonier2002correlations, pal2018quasistatic} and BaTiO$_{3 - x}$H$_x$~\cite{ito2017excited} at temperatures between 100 K and 250 K. The ZF spectra for a sample exhibiting muon diffusion would see slower depolarization, manifested in an elevated tail at long timescales. This phenomenon is seen at 300 K in the x = 25\% doped sample where the tail at long time depolarizes at a much slower rate. Below this temperature, we see no evidence of muon diffusion and the change in the shape of the ZF spectra follows the pattern observed for the other doping levels.

\begin{figure}[htbp]
    \centering
	\includegraphics[width=\textwidth]{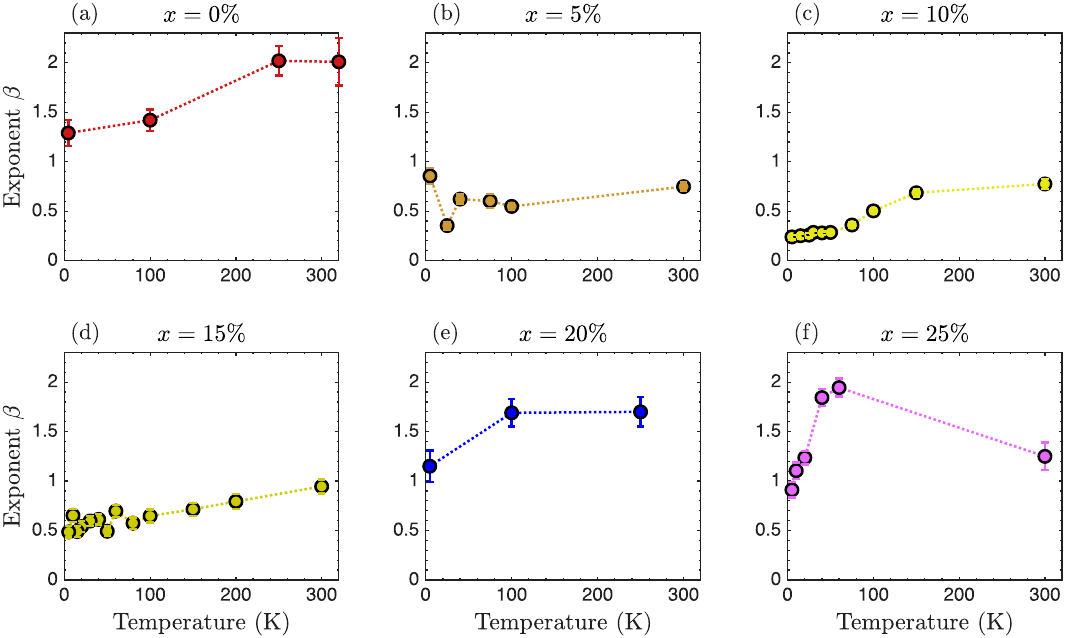}
	\caption{The temperature-dependent lineshape factor, $\beta$, for the nickelate doping series; extracted as a fit parameter using a stretched exponential function describing the ZF spectra. Generally the lineshape factor decreases with temperature, reflecting the continuous change in the spectra shape. The high-temperature $\beta$ extracted for the x = 0.25 sample also indicates muon diffusion.}
\label{fig:Supp_Beta}
\end{figure}

As mentioned in the main text, the ZF spectra can be fitted using a stretched exponential model which may serve as a tool for capturing general trends in the data, specifically the lineshape of the data. The stretched exponential model is given by
\begin{align}
    A^{\rm ZF}
    = A^{\rm ZF}_0e^{-(\lambda_{\rm ZF}  t)^\beta}
\end{align}
where $A_0$ is the initial asymmetry, $\beta$ is the lineshape parameter, $\lambda_{\rm ZF}$ is the depolarization rate, and $t$ is time. The most important parameters is the the lineshape factor, $\beta$, which determines the shape of the asymmetry curve. For example, the Gaussian Kubo-Toyabe function, characterizing only nuclear moments, would have a lineshape parameter of $\beta$ = 2. As static magnetism develops, increased muon depolarization leads to stronger damping of the asymmetry and a reduction of $\beta$ from the Kubo-Toyabe limit ($\beta$ = 2). In the nickelate doping series, $\beta$ decreases systematically with decreasing temperature. This behavior closely tracks the evolution of the ZF spectra, reflecting the corresponding changes in the magnetic state over the measured temperature range.

\section{C. Longitudinal Field fitting function}
The solid red lines shown in Figure~\ref{fig:LF} represent a fit curve of the dynamic gaussian Kubo-Toyabe function. The fitting function is shown below

\begin{align}
A_{\rm const} + \frac{1}{2\pi i}
\int_{\nu-i\infty}^{\nu+i\infty}
\frac{f_G(s+\nu)}{1-B_{ext} f_G(s+\nu)}
\, e^{s t}\, ds
&:= G^{G}_{\mathrm{dyn}}(B_{ext},\sigma,\nu,t),
\\[1ex]
\text{where}\quad
f_G(s)
&=
\int_{0}^{\infty}
G^{\mathrm{LF}}_{G}(t)\, e^{-s t}\, dt .
\end{align}
where $G^{G}_{\mathrm{dyn}}(B_{ext},\sigma,\nu,t)$ describes the time-dependent muon polarization within a Gaussian distribution of fluctuating local fields, while in the presence of an applied longitudinal field. The fit function includes the external field (B$_{ext}$), time after muon implantation ($t$), the fluctuation rate ($\nu$), and the static Gaussian field width ($\sigma$). The fluctuation rate and field width are extracted parameters from the fit function. The fit function also includes $G^{\mathrm{LF}}_{G}(t)$ which is the static LF fit function. The equation takes the Laplace transform of the static function and moves into the frequency domain to more easily account for dynamics under the strong collision model.

% \section{D. Atomic force microscopy characterization}

% Figure~\ref{fig:Supp_AFM} shows a 10 $\mu$m x 10 $\mu$m AFM scan of a representative x = 5\% sample. The sample surface is generally flat with a root-mean-squared roughness of 5 \AA~ and visible step terraces. We see a very small concentration of surface particles over the scanning region with an upper limit of 0.9 inclusions per $\mu$m$^2$. 

% \begin{figure}[htbp]
%     \centering
% 	\includegraphics[width=0.50\textwidth]{supplementary/Figures/LSNO_AFM.png}
% 	\caption{A 10 $\mu$m x 10 $\mu$m AFM topography image of a representative nickelate sample, the present image with x = 5\% Sr doping. The surface is mostly flat with few nm-scale particulates and step terraces visible.}
% \label{fig:Supp_AFM}
% \end{figure}

% We note that previous work using scanning SQUID microscopy on superconducting infinite-layer nickelate films have identified a ferromagnetic background attributed to NiO$_x$ particles found at the interface between the nickelate and SrTiO$_3$ capping layer~\cite{shi2024scanning}. The surface particles were identified using atomic force microscopy (AFM) as 6 nm tall protrusion with a lateral density of 37.5 particles per $\mu$m$^2$. If the observed, sparse particles on our sample surface are indeed NiO$_x$, their contribution to the magnetic properties are expected to be at least one order of magnitude smaller than in previous generations of infinite layer nickelate samples.

\section{D. Run Log}
Tables~\ref{tab:ZF_samples}-~\ref{tab:samples_LF} contain a comprehensive list of all the low-energy muon (LEM) measurements. Table~\ref{tab:ZF_samples} shows $\mu$SR experiments performed in zero field (ZF) as a function of temperature for all doping levels. Tables~\ref{tab:wTF_I}-~\ref{tab:wTF_III} list weak transverse field (wTF) measurements as a function of temperature, and specifically for the x =5\% sample over the two measured fields. Lastly, Table~\ref{tab:samples_LF} list longitudinal field measurements for the temperatures and fields used during the experiment. For clarity the measurements are also indexed with the conditions. The run number and year can be  used to call the asymmetry histogram files on musruser.psi.ch.

\begin{table*}[htbp]
\centering
\caption{$\mu$SR run log index for zero field (ZF) asymmetry histograms.}
\label{tab:ZF_samples}
\begin{tabular}{|c|c|c|c|c|c|}
\hline
Sample & T (K) & E (keV) & B (mT) & Run no. & Year \\
\hline
\multirow{4}{*}{x = 0\%}
 & 4.5 & 2.25 & 0 & 4727 & 2021 \\
 & 100 & 2.25 & 0 & 4787 & 2021 \\
 & 250 & 2.25 & 0 & 4747 & 2021 \\
 & 320 & 2.25 & 0 & 4789 & 2021 \\

 \hline
\multirow{6}{*}{x = 5\%}
 & 5   & 3  & 0 & 3508 & 2025 \\
 & 25  & 3  & 0 & 4509 & 2025 \\
 & 40  & 3  & 0 & 3507 & 2025 \\
 & 75  & 3  & 0 & 3510 & 2025 \\
 & 100 & 3  & 0 & 3511 & 2025 \\
 & 300 & 3  & 0 & 3506 & 2025 \\

\hline
\multirow{10}{*}{x = 10\%}
 & 5   & 3 & 0 & 5654 & 2024 \\
 & 15  & 3 & 0 & 5653 & 2024 \\
 & 25  & 3 & 0 & 5655 & 2024 \\
 & 30  & 3 & 0 & 5658 & 2024 \\
 & 40  & 3 & 0 & 5652 & 2024 \\
 & 50  & 3 & 0 & 5657 & 2024 \\
 & 75  & 3 & 0 & 5651 & 2024 \\ 
 & 100 & 3 & 0 & 5656 & 2024 \\
 & 150 & 3 & 0 & 5650 & 2024 \\
 & 300 & 3 & 0 & 5649 & 2024 \\

\hline
\multirow{13}{*}{x = 15\%}
 & 5   & 3 & 0 & 3825 & 2025 \\
 & 10  & 3 & 0 & 3824 & 2025 \\
 & 15  & 3 & 0 & 3823 & 2025 \\
 & 20  & 3 & 0 & 3822, 3535 & 2025 \\
 & 30  & 3 & 0 & 3821 & 2025 \\
 & 40  & 3 & 0 & 3820, 3534 & 2025 \\
 & 50  & 3 & 0 & 3819 & 2025 \\
 & 60  & 3 & 0 & 3818, 3432 & 2025 \\
 & 80  & 3 & 0 & 3817 & 2025 \\
 & 100 & 3 & 0 & 3816 & 2025 \\
 & 150 & 3 & 0 & 3815 & 2025 \\
 & 200 & 3 & 0 & 3814 & 2025 \\
 & 300 & 3 & 0 & 3812 & 2025 \\

\hline
\multirow{3}{*}{x = 20\%}
 & 4.7 & 2.75 & 0 & 4708 & 2021 \\
 & 100 & 2.75 & 0 & 4785 & 2021 \\
 & 250 & 2.75 & 0 & 4722 & 2021 \\

 \hline
\multirow{6}{*}{x = 25\%}
 & 5   & 3 & 0 & 3551 & 2025 \\
 & 10  & 3 & 0 & 3552 & 2025 \\
 & 20  & 3 & 0 & 3553 & 2025 \\
 & 40  & 3 & 0 & 3554 & 2025 \\
 & 60  & 3 & 0 & 3555 & 2025 \\
 & 300 & 3 & 0 & 4568 & 2025 \\
 
\hline
\end{tabular}
\end{table*}

\begin{table*}[htbp]
\centering
\caption{$\mu$SR run log index for weak transverse field (wTF) asymmetry histograms for x = 0\% and x = 5\%}
\label{tab:wTF_I}
\begin{tabular}{|c|c|c|c|c|c|}
\hline
Sample & T (K) & E (keV) & B (mT) & Run no. & Year \\
\hline
\multirow{1}{*}{x = 0\%}
 & 4.5   & 3  & 10 & 4726 & 2021 \\
 & 8     & 3  & 10 & 4728 & 2021 \\
 & 15    & 3  & 10 & 4729 & 2021 \\
 & 27.5  & 3  & 10 & 4737 & 2021 \\
 & 40    & 3  & 10 & 4738 & 2021 \\
 & 50    & 3  & 10 & 4739 & 2021 \\
 & 75    & 3  & 10 & 4740 & 2021 \\
 & 100   & 3  & 10 & 4741 & 2021 \\
 & 125   & 3  & 10 & 4742 & 2021 \\
 & 150   & 3  & 10 & 4743 & 2021 \\
 & 175   & 3  & 10 & 4744 & 2021 \\
 & 200   & 3  & 10 & 4725 & 2021 \\
 & 225   & 3  & 10 & 4745 & 2021 \\
 & 250   & 3  & 10 & 4746 & 2021 \\

\hline
\multirow{1}{*}{x = 5\%}
 & 5     & 3  & 10 & 3503 & 2025 \\
 & 10    & 3  & 10 & 3502 & 2025 \\
 & 20     & 3 & 10 & 3501 & 2025 \\
 & 30     & 3 & 10 & 3500 & 2025 \\
 & 40    & 3  & 10 & 3499 & 2025 \\
 & 45    & 3  & 10 & 3504 & 2025 \\
 & 50    & 3  & 10 & 3498 & 2025 \\
 & 62.5  & 3  & 10 & 3505 & 2025 \\
 & 75    & 3  & 10 & 3597 & 2025 \\
 & 100   & 3  & 10 & 3496 & 2025 \\
 & 150   & 3  & 10 & 3495 & 2025 \\
 & 200   & 3  & 10 & 3494 & 2025 \\
 & 250   & 3  & 10 & 3493 & 2025 \\
 & 300   & 3  & 10 & 3492 & 2025 \\
 & 5     & 3  & 125 & 3523 & 2025 \\
 & 10    & 3  & 125 & 3522 & 2025 \\
 & 20     & 3 & 125 & 3521 & 2025 \\
 & 30     & 3 & 125 & 3520 & 2025 \\
 & 40     & 3 & 125 & 3519 & 2025 \\
 & 50    & 3  & 125 & 3518 & 2025 \\
 & 75    & 3  & 125 & 3517 & 2025 \\
 & 100   & 3  & 125 & 3516 & 2025 \\
 & 150   & 3  & 125 & 3515 & 2025 \\
 & 200   & 3  & 125 & 3514 & 2025 \\
 & 250   & 3  & 125 & 3513 & 2025 \\
 & 300   & 3  & 125 & 3512 & 2025 \\

\hline
\end{tabular}
\end{table*}

\begin{table*}[htbp]
\centering
\caption{$\mu$SR run log index for weak transverse field asymmetry (wTF) histograms for x = 10\%, 15\%, and 20\%}
\label{tab:samples_wTF_II}
\begin{tabular}{|c|c|c|c|c|c|}
\hline
Sample & T (K) & E (keV) & B (mT) & Run no. & Year \\

\hline
\multirow{14}{*}{x = 10\%}
 & 5     & 3 & 0 & 5632 & 2024 \\
 & 10    & 3 & 0 & 5631 & 2024 \\
 & 15    & 3 & 0 & 5633 & 2024 \\
 & 20    & 3 & 0 & 5630 & 2024 \\
 & 30    & 3 & 0 & 5639 & 2024 \\
 & 40    & 3 & 0 & 5629 & 2024 \\
 & 50    & 3 & 0 & 5640 & 2024 \\
 & 75    & 3 & 0 & 5628 & 2024 \\
 & 100   & 3 & 0 & 5641 & 2024 \\
 & 150   & 3 & 0 & 5627 & 2024 \\
 & 200   & 3 & 0 & 5626 & 2024 \\
 & 250   & 3 & 0 & 5642 & 2024 \\
 & 300   & 3 & 0 & 5642 & 2024 \\
 & 320   & 3 & 0 & 5667 & 2024 \\
 
\hline
\multirow{12}{*}{x = 15\%}
 & 5     & 3 & 10 & 3679, (3826) & 2024, (2025) \\
 & 10    & 3 & 10 & 3678, (3859) & 2024, (2025) \\
 & 15    & 3 & 10 & 3858 & 2025 \\
 & 20    & 3 & 10 & 3677, (3857) & 2024, (2025) \\
 & 30    & 3 & 10 & 3676, (3856) & 2024, (2025) \\
 & 40    & 3 & 10 & 3675, (3855) & 2024, (2025) \\
 & 50    & 3 & 10 & 3674, (3854) & 2024, (2025) \\
 & 60    & 3 & 10 & 3853 & 2025 \\
 & 75    & 3 & 10 & 3673 & 2024 \\
 & 80    & 3 & 10 & 3852 & 2025 \\
 & 100   & 3 & 10 & 3672, (3851) & 2024, (2025) \\
 & 150   & 3 & 10 & 3671, (3850) & 2024  (2025) \\
 & 200   & 3 & 10 & 3670, (3849) & 2024, (2025) \\
 & 250   & 3 & 10 & 3680 & 2024 \\
 & 300   & 3 & 10 & 3681, (3848) & 2024, (2025) \\

\hline
\multirow{11}{*}{x = 20\%}
 & 4.7 & 2.75 & 10 & 4707 & 2021 \\
 & 15 & 2.75  & 10 & 4709 & 2021 \\
 & 20 & 2.75  & 10 & 4706 & 2021 \\
 & 30  & 2.75 & 10 & 4715 & 2021 \\
 & 40  & 2.75 & 10 & 4716 & 2021 \\
 & 50  & 2.75 & 10 & 4717 & 2021 \\
 & 75  & 2.75 & 10 & 4718 & 2021 \\
 & 100 & 2.75 & 10 & 4719 & 2021 \\
 & 150 & 2.75 & 10 & 4720 & 2021 \\
 & 200 & 2.75 & 10 & 4705 & 2021 \\
 & 250 & 2.75 & 10 & 4721 & 2021 \\

% \hline
% \multirow{12}{*}{x = 25\%}
%  & 5    & 3 & 10 & 3690 & 2024 \\
%  & 7.5  & 3 & 10 & 3558 & 2025 \\
%  & 10   & 3 & 10 & 3691 & 2024 \\
%  & 12.5 & 3 & 10 & 3559 & 2025 \\
%  & 15   & 3 & 10 & 3560 & 2025 \\
%  & 17.5 & 3 & 10 & 3561 & 2025 \\
%  & 20   & 3 & 10 & 3692, (3562) & 2024, (2025) \\
%  & 30   & 3 & 10 & 3693, (3563) & 2024, (2025) \\
%  & 40   & 3 & 10 & 3694 & 2024 \\
%  & 50   & 3 & 10 & 3695, (3565) & 2024, (2025) \\
%  & 75   & 3 & 10 & 3696 & 2024 \\
%  & 100  & 3 & 10 & 3697, (3565) & 2024, (2025) \\
%  & 150  & 3 & 10 & 3698 & 2024 \\
%  & 200  & 3 & 10 & 3699, (3566) & 2024, (2025) \\
%  & 250  & 3 & 10 & 3700 & 2024 \\
%  & 300  & 3 & 100 & 3701, (3567) & 2024, (2025) \\

\hline
\end{tabular}
\end{table*}

\begin{table*}[htbp]
\centering
\caption{$\mu$SR run log index for weak transverse field (wTF) asymmetry histograms for x = 25\%}
\label{tab:wTF_III}
\begin{tabular}{|c|c|c|c|c|c|}
\hline
Sample & T (K) & E (keV) & B (mT) & Run no. & Year \\
\hline
\multirow{12}{*}{x = 25\%}
 & 5    & 3 & 10 & 3690 & 2024 \\
 & 7.5  & 3 & 10 & 3558 & 2025 \\
 & 10   & 3 & 10 & 3691 & 2024 \\
 & 12.5 & 3 & 10 & 3559 & 2025 \\
 & 15   & 3 & 10 & 3560 & 2025 \\
 & 17.5 & 3 & 10 & 3561 & 2025 \\
 & 20   & 3 & 10 & 3692, (3562) & 2024, (2025) \\
 & 30   & 3 & 10 & 3693, (3563) & 2024, (2025) \\
 & 40   & 3 & 10 & 3694 & 2024 \\
 & 50   & 3 & 10 & 3695, (3565) & 2024, (2025) \\
 & 75   & 3 & 10 & 3696 & 2024 \\
 & 100  & 3 & 10 & 3697, (3565) & 2024, (2025) \\
 & 150  & 3 & 10 & 3698 & 2024 \\
 & 200  & 3 & 10 & 3699, (3566) & 2024, (2025) \\
 & 250  & 3 & 10 & 3700 & 2024 \\
 & 300  & 3 & 100 & 3701, (3567) & 2024, (2025) \\

\hline
\end{tabular}
\end{table*}

\begin{table}[htbp]
\centering
\caption{$\mu$SR run log index for longitudinal field (LF) asymmetry histograms.}
\label{tab:samples_LF}
\begin{tabular}{|c|c|c|c|c|c|}
\hline
Sample & T (K) & E (keV) & B (G) & Run no. & Year \\

\hline
\multirow{4}{*}{x = 10\%}
 & 5   & 3 & 25  & 5659, 5660 & 2024 \\
 & 5   & 3 & 100 & 5661, 5662 & 2024 \\
 & 45  & 3 & 25  & 5663, 5664 & 2024 \\
 & 45  & 3 & 100 & 5665, 5666 & 2024 \\

\hline
\multirow{4}{*}{x = 15\%}
 & 5   & 3 & 25   & 3528, 3529 & 2025 \\
 & 5   & 3 & 100  & 3530, 3531 & 2025 \\
 & 40  & 3 & 25   & 3524, 3525 & 2025 \\
 & 40  & 3 & 100  & 3526, 3527 & 2025 \\

\hline
\multirow{4}{*}{x = 25\%}
 & 5   & 3 & 25   & 3547, 3548 & 2025 \\
 & 5   & 3 & 100  & 3549, 3550 & 2025 \\
 & 40  & 3 & 25   & 3543, 3544 & 2025 \\
 & 40  & 3 & 100  & 3545, 3546 & 2025 \\
 
\hline
\end{tabular}
\end{table}

\end{document}